# Multi-Issue Negotiation with Deadlines

**Shaheen S. Fatima**                                    S.S.FATIMA@CSC.LIV.AC.UK
**Michael Wooldridge**                                   M.J.WOOLDRIDGE@CSC.LIV.AC.UK
*Department of Computer Science,*
*University of Liverpool, Liverpool L69 3BX, U.K.*

**Nicholas R. Jennings**                                 NRJ@ECS.SOTON.AC.UK
*School of Electronics and Computer Science,*
*University of Southampton, Southampton SO17 1BJ, U.K.*

## Abstract

This paper studies bilateral multi-issue negotiation between self-interested autonomous agents. Now, there are a number of different procedures that can be used for this process; the three main ones being the *package deal procedure* in which all the issues are bundled and discussed together, the *simultaneous procedure* in which the issues are discussed simultaneously but independently of each other, and the *sequential procedure* in which the issues are discussed one after another. Since each of them yields a different outcome, a key problem is to decide which one to use in which circumstances. Specifically, we consider this question for a model in which the agents have time constraints (in the form of both deadlines and discount factors) and information uncertainty (in that the agents do not know the opponent's utility function). For this model, we consider issues that are both independent and those that are interdependent and determine equilibria for each case for each procedure. In so doing, we show that the package deal is in fact the optimal procedure for each party. We then go on to show that, although the package deal may be computationally more complex than the other two procedures, it generates Pareto optimal outcomes (unlike the other two), it has similar earliest and latest possible times of agreement to the simultaneous procedure (which is better than the sequential procedure), and that it (like the other two procedures) generates a unique outcome only under certain conditions (which we define).

## 1. Introduction

Negotiation is a key form of interaction in multiagent systems (Maes, Guttman, & Moukas, 1999; Sandholm, 2000). It is a process in which disputing agents decide how to divide the gains from cooperation. Since this decision is made jointly by the agents themselves (Rosenschein & Zlotkin, 1994; Raiffa, 1982; Pruitt, 1981; Fisher & Ury, 1981; Young, 1975; Kraus, 2001), each agent can only obtain what the other is prepared to allow them. Now, the simplest form of negotiation involves two agents and a single-issue. For example, consider a scenario in which a buyer and a seller negotiate on the price of a good. To begin, the two agents are likely to differ on the price at which they believe the trade should take place, but through a process of joint decision-making they either arrive at a price that is mutually acceptable or they fail to reach an agreement. Since agents are likely to begin with different prices, one or both of them must move toward the other, through a series of offers and counter offers, in order to obtain a mutually acceptable outcome. However, before the agents can actually perform such negotiations, they must decide the rules for making offers and counter offers. That is, they must set the negotiation protocol (Lax & Sebenius, 1986; Osborne & Rubinstein, 1990; Rosenschein & Zlotkin, 1994; Kraus, Wilkenfeld, & Zlotkin, 1995; Lomuscio, Wooldridge, & Jennings, 2003).

On the basis of this protocol, each agent chooses its strategy (i.e., what offers it should make during the course of negotiation). For competitive scenarios with self-interested agents, each participant defines its strategy so as to maximise its individual utility. Furthermore, for such scenarios, an agent's optimal strategy depends very strongly on the information it has about its opponent (Fatima, Wooldridge, & Jennings, 2002, 2004). For example, the strategy that a buyer would use if it knew the seller's reserve price differs from the one it would use if it did not. From all of this, it can be seen that the outcome of single-issue negotiation depends on four key factors (Harsanyi, 1977): the negotiation protocol, the players' strategies, the players' preferences over the possible outcomes, and the information that the players have about each other. However, in most bilateral negotiations, the parties involved need to settle more than one issue. For example, agents may need to come to agreements about objects/services that are characterised by attributes such as price, delivery time, quality, reliability, and so on. For such multi-issue negotiations, the outcome also depends on one additional factor: the *negotiation procedure* (Schelling, 1956, 1960; Fershtman, 1990), which specifies how the issues will be settled. Broadly speaking, there are three ways of negotiating multiple issues (Keeney & Raiffa, 1976; Raiffa, 1982):

- *Package deal*: This approach links all the issues and discusses them together as bundle.

- *Simultaneous negotiation*: This involves settling the issues simultaneously, but independently, of each other.

- *Sequential negotiation*: This involves negotiating the issues sequentially, one after another.

Now, these three different procedures have different properties and yield different outcomes to the negotiators (Fershtman, 2000). So the key question to answer is: which of them is best? Here, since we are concerned with self-interested agents, our notion of the optimal procedure is the one that maximises an agent's individual return. However, such optimality is only part of the story; given our motivations we are also concerned with the Pareto optimality of the solutions for these procedures (because Pareto optimality ensures that utility does not go wasted), the computational complexity of the procedures (because for scenarios with information uncertainty, the agents need to compute their equilibrium offers during the process of negotiation, as opposed to the complete information scenario where the strategies can be precompiled), the actual time of agreement (because for scenarios with information uncertainty, this time depends on an agent's beliefs about its opponent and an agreement may not occur in the first time period), and the uniqueness of the solutions they generate (because this allows the agents to know their actual shares).

One immediate observation in this vein is that the package deal gives rise to the possibility of making tradeoffs across issues. Such tradeoffs are possible when different agents value different issues differently. For example, if there are two issues and one agent values the first more than the second, while the other agent values the second issue more than the first, then it is possible to make tradeoffs and thereby improve the utility of both agents relative to the situation without tradeoffs. In contrast, for the simultaneous and sequential approaches, the issues are settled independently and so there is no scope for such tradeoffs between them. Moreover, we seek to answer the above question about optimality for the types of situation that are commonly faced by agents in real-world contexts. Thus, we consider negotiations in which there are:

1. *Time constraints.* Agents have time constraints in the form of both deadlines and discount factors. Here we view deadlines as an essential element since negotiation cannot go on indefinitely, rather it must end within a reasonable time limit (Livne, 1979). Likewise, discount factors are essential since the desirability of the good being traded often declines with time. This happens either because the good is perishable or due to inflation. Moreover, the strategic behaviour of agents with deadlines and discount factors differs from those without (see Rubinstein, 1982, for single issue bargaining without deadlines and Sandholm & Vulkan, 1999; Ma & Manove, 1993; Fershtman & Seidmann, 1993; Kraus, 2001, for bargaining with deadlines and discount factors). For instance, the presence of a deadline induces each negotiator to play a strategy that ensures the best possible agreement before the deadline is reached. Likewise, the presence of a discount factor means that reaching an agreement today is not the same as reaching it tomorrow. Hence, the agents try to reach an agreement sooner rather than later.

2. *Uncertainty about the opponent's negotiation parameters.* The information that agents have about their negotiation opponent is likely to be uncertain (see Fudenberg & Tirole, 1983; Fudenberg, Levine, & Tirole, 1985; Rubinstein, 1985, for single issue bargaining with uncertainty). Moreover, in some bargaining situations, one of the players may know something of relevance that the other does not. For example, when bargaining over the price of a second hand car, the seller knows its quality, but the buyer does not. Such situations are said to have *asymmetry* in information between the players (Muthoo, 1999). On the other hand, in *symmetric* information situations both players have the same information. Again, agents have to operate in both situations and so we analyse both cases.

3. *Interdependence between issues.* The issues under negotiation may be independent or interdependent. In the former case, an agent's utility from an issue depends only on the agreement that is reached on it, not on how the other issues are settled. In the latter case, an agent's utility from an issue depends not only on the agreement that is reached on it but also on how the other issues are settled (Bar-Yam, 1997; Klein, Faratin, Sayama, & Bar-Yam, 2003). Both situations are common in multiagent systems and so again we analyse both cases.

Thus we study five different settings: i) complete information setting ($CI$), ii) a setting with independent issues and symmetric uncertainty about the agents' utilities ($SU_I$), iii) a setting with independent issues and asymmetric uncertainty about the agents' utilities ($AU_I$), iv) a setting with interdependent issues and symmetric uncertainty about the agents' utilities ($SU_D$), and v) a setting with interdependent issues and asymmetric uncertainty about the agents' utilities ($AU_D$).

Our methodology is to first derive equilibria for each of the procedures in each of the above settings, From this, we can determine which of them is optimal. As we will see, this analysis shows that, for all the settings, the package deal is the best. We then go on to analyse the procedures in terms of other performance metrics. Specifically, we show that, in all the settings, only the package deal generates a Pareto optimal outcome. We also show that although the package deal may be computationally more complex than the other two procedures, it has similar earliest and latest possible times of agreement to the simultaneous procedure (which is better than the sequential procedure), and it (like the other two procedures) generates a unique outcome only in certain situations (which we define). The key results of our study are summarised in Table 1.

There has previously been some formal comparison of different procedures to find the optimal one (see Section 7 for details). However, all this work has at least one of the following major limitations. First, it has focused on comparing procedures for negotiation without deadlines. Note that existing work has obtained equilibrium for negotiation with deadlines, but only for the single issue case (Sandholm & Vulkan, 1999; Stahl, 1972), and a special type of the sequential procedure for multiple issues (Fatima et al., 2004). See Section 7 for details. Second, it has focussed only on independent issues and asymmetric information settings. Third, it has only focused on finding the optimal procedure, but has not considered the additional solution properties of different procedures. Given this, our paper makes a threefold contribution. First, we obtain the equilibrium for each procedure when there are deadlines. Second, we analyse multiple issues that are both independent and interdependent. Moreover, we analyse both symmetric and asymmetric information settings. Finally, on the basis of the equilibrium for different procedures, we provide the first comprehensive comparison of their solution properties (viz. time complexity, Pareto optimality, uniqueness, and time of agreement). When taken together, the results clearly indicate the choices and tradeoffs involved in choosing a negotiation procedure in a wide range of circumstances. This knowledge can be used by a system designer who is responsible for designing the mechanism that should be used to moderate the negotiation encounters and by the agents themselves if they can choose how to arrange their interactions. Furthermore, this knowledge also tells the agents what their equilibrium offers are during negotiation.

| | Information setting | Package deal | Simultaneous | Sequential |
|---|---|---|---|---|
| Time of agreement $t_c$ | $CI$ | For the $c$th issue $t_c = 1$ for $1 \leq c \leq m$ | For the $c$th issue $t_c = 1$ for $1 \leq c \leq m$ | For the $c$th partition $t_c = c$ for $1 \leq c \leq \mu$ |
| | $SU_I, SU_D$ $AU_I$, and $AU_D$ | For the $c$th issue $t_c^e = 1$ $t_c^l = min(2r-1, n)$ for $1 \leq c \leq m$ | For the $c$th issue $t_c^e = 1$ $t_c^l = min(2r-1, n)$ for $1 \leq c \leq m$ | For the $c$th partition $t_c^e = t_c^s$ $t_c^l = t_c^s + min(2r-1, n)$ for $1 \leq c \leq \mu$ |
| Time to compute equilibrium | $CI$ | $\mathcal{O}(mn)$ | $\mathcal{O}(Mn)$ | $\mathcal{O}(Mn)$ |
| | $SU_I$ and $SU_D$ | $\mathcal{O}(m\hat{\pi}r^3T(n-\frac{T}{2}))$ | $\mathcal{O}(|S_z|\hat{\pi}_z r^3 T(n-\frac{T}{2}))$ | $\mathcal{O}(|S_z|\hat{\pi}_z r^3 T(n-\frac{T}{2}))$ |
| | $AU_I$ and $AU_D$ | $\mathcal{O}(m\hat{\pi}r^3(n-\frac{T}{2})\frac{T}{2})$ | $\mathcal{O}(|S_z|\hat{\pi}_z r^3(n-\frac{T}{2})\frac{T}{2})$ | $\mathcal{O}(|S_z|\hat{\pi}_z r^3(n-\frac{T}{2})\frac{T}{2})$ |
| Pareto optimal? | $CI$, $SU_I, SU_D$, $AU_I$, and $AU_D$ | Yes | No | No |
| Unique equilibrium? | $CI$ | If $\neg C_1$ | If $C_2$ | If $C_2$ |
| | $SU_I, SU_D$, $AU_I$, and $AU_D$ | If $\neg C_3 \vee C_4$ | If $C_5$ | If $C_5$ |

Table 1: A summary of key results. $t_c^s$ denotes the start time for the $c$th partition, $t_c^e$ the earliest possible time of agreement, and $t_c^l$ the latest possible time of agreement).

The remainder of the paper is organised as follows. We begin by giving a brief overview of single-issue negotiation in Section 2. In Section 3, we study the three multi-issue procedures for the setting with complete information and where the issues are independent. This study is undertaken to provide a foundation for Sections 4, 5, and 6, which treat the information about the agents' utilities as uncertain. More specifically, in Section 4, we analyse a scenario with symmetric uncer-tainty about the opponent's utility. In Section 5, we analyse a scenario with asymmetric uncertainty about the opponent's utility. Sections 4 and 5 both deal with independent issues. In Section 6, we extend the analysis to interdependent issues. Section 7 discusses the related literature and Section 8 concludes. Appendix A provides a summary of notation employed throughout the paper.

## 2. Single-Issue Negotiation

Assume there are two agents: $a$ and $b$. Each agent has time constraints in the form of deadlines and discount factors. Since we focus on competitive scenarios with self-interested agents, we model negotiation using the 'split the pie game' analysed by Osborne and Rubinstein (1994), Binmore, Osborne, and Rubinstein (1992). We begin by introducing this complete information game.

Let the two agents be negotiating over a single issue ($i$). This issue is a 'pie' of size 1 and the agents want to determine how to divide it between themselves. There is a deadline (i.e., a number of rounds by which negotiation must end). Let $n \in \mathbb{N}^+$ denote this deadline. The agents use Rubinstein's alternating offers protocol (Osborne & Rubinstein, 1994), which proceeds through a series of time periods. One of the agents, say $a$, starts negotiation in the first time period (i.e., $t = 1$) by making an offer ($x_i$), that lies in the interval $[0, 1]$, to $b$. Agent $b$ can either accept or reject the offer. If it accepts, negotiation ends in an agreement with $a$ getting a share of $x_i$ and $b$ getting $y_i = 1 - x_i$. Otherwise, negotiation proceeds to the next time period, in which agent $b$ makes a counter-offer. This process of making offers continues until one of the agents either accepts an offer or quits negotiation (resulting in a conflict). Thus, there are three possible actions an agent can take during any time period: accept the last offer, make a new counter-offer, or quit the negotiation.

An essential feature of negotiations involving alternating offers is that the pie is assumed to shrink with time (Rubinstein, 1982). Specifically, it shrinks at each step of offer and counteroffer. This shrinkage models a decrease in the value of the pie (representing the fact that the pie perishes with time or there is inflation). This shrinkage is represented with a discount factor denoted $0 < \delta_i \leq 1$ for both[1] agents. At $t = 1$, the size of the pie is 1, but in all subsequent time periods $t > 1$, the pie shrinks to $\delta_i^{t-1}$.

We denote the set of real numbers by $\mathbb{R}$ and the set of real numbers in the interval $[0, 1]$ by $\mathbb{R}_1$. Then let $[x_i^t, y_i^t]$ denote the offer made at time period $t$ where $x_i^t$ and $y_i^t$ denote the share for agent $a$ and $b$ respectively. Then, for a given pie, the set of possible offers is:

$$\{[x_i^t, y_i^t] : x_i^t \geq 0, \ y_i^t \geq 0, \ and \ x_i^t + y_i^t = \delta_i^{t-1}\}$$

where $x_i^t \in \mathbb{R}_1$ and $y_i^t \in \mathbb{R}_1$. Each player's utility function is defined over the set $\mathbb{R}$. Let $u_i^a : \mathbb{R}_1 \times \mathbb{N}^+ \to \mathbb{R}$ and $u_i^b : \mathbb{R}_1 \times \mathbb{N}^+ \to \mathbb{R}$ denote the utility functions of the two agents. At time $t$, if $a$ and $b$ receive a share of $x_i^t$ and $y_i^t$ respectively (where $x_i^t + y_i^t = \delta_i^{t-1}$), then their utilities are:

$$u_i^a(x_i^t, t) = \begin{cases} x_i^t & \text{if } t \leq n \\ 0 & \text{otherwise} \end{cases}$$

$$u_i^b(y_i^t, t) = \begin{cases} y_i^t & \text{if } t \leq n \\ 0 & \text{otherwise} \end{cases}$$

---

1. Having a different discount factor for different agents only makes the presentation more involved without leading to any changes in the analysis of the strategic behaviour of the agents or the time complexity of finding the equilibrium offers. Hence we have a single discount factor for both agents.

The conflict utility (i.e., the utility received in the event that no deal is struck) is zero for both agents. Note that $\delta$ is not shown explicitly in an agent's utility function but is implicit. This is because, during any time period $t$, $x_i^t$ and $y_i^t$ denote $a$'s and $b$'s actual shares respectively (not the ratios of their shares) where $x_i^t + y_i^t = \delta_i^{t-1}$. In other words $\delta$ is included in an agent's share. This will become clearer when we show the agents' shares in Expression 1.

For the above setting, the agents reason as follows in order to determine what to offer. Let agent $a$ denote the first mover (i.e., at $t = 1$, $a$ proposes to $b$ how to split the pie). To begin, consider the case where the deadline for both agents is $n = 1$. If $b$ accepts, the division occurs as agreed; if not, neither agent gets anything (since $n = 1$ is the deadline). Here, $a$ is in a powerful position and is able to propose to keep 100 percent of the pie and give nothing to $b$ [2]. Since the deadline is $n = 1$, $b$ accepts this offer and agreement takes place in the first time period.

Now, consider the case where the deadline is $n = 2$. In the first round, the size of the pie is 1 but it shrinks to $\delta_i$ in the second round. In order to decide what to offer in the first round, $a$ looks ahead to $t = 2$ and reasons backwards [3]. Agent $a$ reasons that if negotiation proceeds to the second round, $b$ will take 100 percent of the shrunken pie by offering $[0, \delta_i]$ and leave nothing for $a$. Thus, in the first time period, if $a$ offers $b$ anything less than $\delta_i$, $b$ will reject the offer. Hence, during the first time period, agent $a$ offers $[1 - \delta_i, \delta_i]$. Agent $b$ accepts this and an agreement occurs in the first time period.

In general, if the deadline is $n$, negotiation proceeds as follows. As before, agent $a$ decides what to offer in the first round by looking ahead as far as $t = n$ and then reasoning backwards. This decision making leads $a$ to make the following offer in the first time period:

$$[\Sigma_{j=0}^{n-1}[(-1)^j \delta_i^j], 1 - \Sigma_{j=0}^{n-1}[(-1)^j \delta_i^j]] \tag{1}$$

Agent $b$ accepts this offer and negotiation ends in the first time period. Note that the equilibrium outcome depends on who makes the first move. Since we have two agents and either of them could move first, we get two possible equilibrium outcomes.

On the basis of the above equilibrium for single-issue negotiation with complete information, we first obtain the equilibrium for multiple issues and then determine the optimal negotiation procedure for the various settings that we have previously described.

## 3. Multi-Issue Negotiation with Complete Information

As mentioned in Section 1, the existing literature does not provide an analysis of all the multi-issue procedures for negotiation with deadlines. Hence, we begin by analysing the complete information setting. From this base, we can then extend to the case where there is information uncertainty.

Here $a$ and $b$ negotiate over $m > 1$ independent issues (Section 6 deals with interdependent issues). These issues are $m$ distinct pies and the agents want to determine how to split each of them. Let $S = \{1, 2, \ldots, m\}$ denote the set of $m$ pies. As before, each pie is of size 1. Let the discount factor for issue $c$, where $1 \le c \le m$, be $0 < \delta_c \le 1$. For each issue, let $n$ denote each agent's deadline. In the offer for time period $t$ (where $1 \leq t \leq n$), agent $a$'s ($b$'s) share for each of the $m$ issues is now represented as an $m$ element vector $x^t \in \mathbb{R}_1^m$ ($y^t \in \mathbb{R}_1^m$). Thus, if agent $a$'s share for issue $c$ at time $t$ is $x_c^t$, then agent $b$'s share is $y_c^t = (\delta_c^{t-1} - x_c^t)$. The shares for $a$ and $b$ are together represented as the package $[x^t, y^t]$.

---

2. It is possible that $b$ may reject such a proposal. In practice, $a$ will have to propose an offer that is just enough to induce $b$ to accept. However, to keep the exposition simple, we assume that $a$ can get the whole pie by making the 100 percent proposal.

3. This backward reasoning method is adopted from (Stahl, 1972). Our model is a generalisation of (Stahl, 1972); during time period $t$, an agent in our model can propose any offer between zero and $\delta^{t-1}$ (because the size of the pie is $\delta^{t-1}$), but a player in (Stahl, 1972) is given a fixed number of alternatives to choose from.

We define an agent's cumulative utility using the additive form. There are two reasons for this. First, it is the most common form for cumulative utilities in traditional multi-issue utility theory (Keeney & Raiffa, 1976). Second, additive cumulative utilities are linear and so the problem of making tradeoffs becomes computationally tractable[4]. The functions $U^a : \mathbb{R}_1^m \times \mathbb{R}_1^m \times \mathbb{N}^+ \to \mathbb{R}$ and $U^b : \mathbb{R}_1^m \times \mathbb{R}_1^m \times \mathbb{N}^+ \to \mathbb{R}$ give the cumulative utilities for $a$ and $b$ respectively at time $t$. These are defined as follows:

$$U^a([x^t, y^t], t) = \begin{cases} \Sigma_{c=1}^m k_c^a u_c^a(x_c^t, t) & \text{if } t \leq n \\ 0 & \text{otherwise} \end{cases} \qquad (2)$$

$$U^b([x^t, y^t], t) = \begin{cases} \Sigma_{c=1}^m k_c^b u_c^b(y_c^t, t) & \text{if } t \leq n \\ 0 & \text{otherwise} \end{cases} \qquad (3)$$

where $k^a \in \mathbb{R}_+^m$ denotes an $m$ element vector of constants for agent $a$ and $k^b \in \mathbb{R}_+^m$ that for $b$. Here $\mathbb{R}_+$ denotes the set of positive real numbers. These vectors indicate how the agents value different issues. For example, if $k_c^a > k_{c+1}^a$, then agent $a$ values issue $c$ more than issue $c + 1$. Likewise for agent $b$. In other words, the $m$ issues are *perfect substitutes* (i.e., all that matters to an agent is its total utility for all the $m$ issues and not that for any subset of those Varian, 2003; Mas-Colell, Whinston, & Green, 1995). In all the settings we study, the issues will be perfect substitutes.

Each agent has complete information about all negotiation parameters (i.e., $n$, $m$, $k_c^a$, $k_c^b$, and $\delta_c$ for $1 \leq c \leq m$). For this complete information setting, we now determine the equilibrium for the package deal, the simultaneous procedure, and the sequential procedure.

## 3.1 The Package Deal Procedure

In this procedure, the agents use the same protocol as for single-issue negotiation (described in Section 2). However, an offer for the package deal includes a proposal for each issue under negotiation. Thus, for $m$ issues, an offer includes $m$ divisions, one for each issue. Agents are allowed to either accept a complete offer (i.e., all $m$ issues) or reject a complete offer. An agreement can therefore take place either on all $m$ issues or on none of them.

As per single-issue negotiation, an agent decides what to offer by looking ahead and reasoning backwards. However, since an offer for the package deal includes a share for all the $m$ issues, agents can now make tradeoffs across the issues in order to maximise their cumulative utilities. For $1 \leq c \leq m$, the equilibrium offer for issue $c$ at time $t$ is denoted as $[a_c^t, b_c^t]$ where $a_c^t$ and $b_c^t$ denote the shares for agent $a$ and $b$ respectively. We denote the equilibrium package at time $t$ as $[a^t, b^t]$

---

4. Using a form other than the additive one will make the function nonlinear. Consequently an agent's tradeoff problem becomes a *global optimization problem* with a *nonlinear* objective function. Due to their computational complexity, such nonlinear optimization problems can only be solved using *approximation methods* (Horst & Tuy, 1996; Bar-Yam, 1997; Klein et al., 2003). Moreover, these methods are not general in that they depend on how the cumulative utilities are actually defined. In order to overcome this difficulty, we used the additive form for defining cumulative utilities. Consequently, our tradeoff problem is a linear optimization problem, the *exact* solution to which can be found in polynomial time (as shown in Theorems 1 and 2). Although our results apply to the above defined additive cumulative utilities, in Section 6.4 we discuss how they would hold for nonlinear utilities.

where $a^t \in \mathbb{R}_1^m$ ($b^t \in \mathbb{R}_1^m$) is an $m$ element vector that denotes $a$'s ($b$'s) share for each of the $m$ issues. Also, for $1 \leq t \leq n$, $\delta^{t-1} \in \mathbb{R}_1^m$ is an $m$ element vector that represents the sizes of the $m$ pies at time $t$. The symbol $\mathbf{0}$ denotes an $m$ element vector of zeroes. Note that for $1 \leq t \leq n$, $a^t + b^t = \delta^{t-1}$ (i.e., the sum of the agents' shares (at time $t$) for each pie is equal to the size of the pie at $t$). Finally, for time period $t$ (for $1 \leq t \leq n$) we let A($t$) (respectively B($t$)) denote the equilibrium strategy for agent $a$ (respectively $b$).

As mentioned in Section 1, the package deal allows agents to make tradeoffs. We let TRADEOFFA (TRADEOFFB) denote agent $a$'s ($b$'s) function for making tradeoffs. Given this, the following theorem characterises the equilibrium for the package deal procedure.

**Theorem 1** *For the package deal procedure, the following strategies form a Nash equilibrium. The equilibrium strategy for $t = n$ is:*

$$\text{A}(n) = \begin{cases} OFFER \ [\delta^{n-1}, \mathbf{0}] & IF\ a\text{'s TURN} \\ ACCEPT & IF\ b\text{'s TURN} \end{cases}$$

$$\text{B}(n) = \begin{cases} OFFER \ [\mathbf{0}, \delta^{n-1}] & IF\ b\text{'s TURN} \\ ACCEPT & IF\ a\text{'s TURN} \end{cases}$$

*For all preceding time periods $t < n$, if $[x^t, y^t]$ denotes the offer made at time $t$, then the equilibrium strategies are defined as follows:*

$$\text{A}(t) = \begin{cases} OFFER \ \text{TRADEOFFA}(k^a, k^b, \delta, \text{UB}(t), m, t) & IF\ a\text{'s TURN} \\ If\ (U^a([x^t, y^t], t) \geq \text{UA}(t))\ ACCEPT\ else\ REJECT & IF\ b\text{'s TURN} \end{cases}$$

$$\text{B}(t) = \begin{cases} OFFER \ \text{TRADEOFFB}(k^a, k^b, \delta, \text{UA}(t), m, t) & IF\ b\text{'s TURN} \\ If\ (U^b([x^t, y^t], t) \geq \text{UB}(t))\ ACCEPT\ else\ REJECT & IF\ a\text{'s TURN} \end{cases}$$

*where $\text{UA}(t) = U^a([a^{t+1}, b^{t+1}], t+1)$ and $\text{UB}(t) = U^b([a^{t+1}, b^{t+1}], t+1)$. An agreement takes place at $t = 1$.*

**Proof:** We look ahead to the last time period (i.e., $t = n$) and then reason backwards. To begin, if negotiation reaches the deadline ($n$), then the agent whose turn it is takes everything and leaves nothing for its opponent. Hence, we get the strategies A($n$) and B($n$) as given in the statement of the theorem.

In all the preceding time periods ($t < n$), the offering agent proposes a package that gives its opponent a cumulative utility equal to what the opponent would get from its own equilibrium offer for the next time period. During time period $t$, either $a$ or $b$ could be the offering agent. Consider the case where $a$ makes an offer at $t$. The package that $a$ offers at $t$ gives $b$ a cumulative utility of $U^b([a^{t+1}, b^{t+1}], t+1)$. However, since there is more than one issue, there is more than one package that gives $b$ a cumulative utility of $U^b([a^{t+1}, b^{t+1}], t+1)$. From among these packages, $a$ offers the one that maximises its own cumulative utility (because it is a utility maximiser). Thus, the problem for $a$ is to find the package $[a^t, b^t]$ so as to:

$$
\begin{aligned}
&\text{maximise} && \Sigma_{c=1}^{m} k_c^a a_c^t \\
&\text{such that} && \Sigma_{c=1}^{m} (\delta_c^{t-1} - a_c^t) k_c^b = \text{UB}(t) \\
&&& 0 \leq a_c^t \leq 1 \qquad \text{for } 1 \leq c \leq m
\end{aligned}
\tag{4}
$$

This tradeoff problem is similar to the *fractional knapsack problem* (Martello & Toth, 1990; Cormen, Leiserson, Rivest, & Stein, 2003), the optimal solution for which can be generated using a *greedy* approach[5] (i.e., by filling the knapsack with items in the decreasing order of value per unit weight). The items in the knapsack problem are analogous to the issues in our case. The only difference is that the fractional knapsack problem starts with an empty knapsack and aims to fill it with items so as to maximise the cumulative value, while an agent's tradeoff problem can be viewed as starting with the agent having 100 per cent of all the issues and then aiming to give away portions of issues to its opponent so that the latter gets a given cumulative utility, while the resulting loss in its own utility is minimised. Thus, in order to find how to split the $m$ issues, agent $a$ considers $k_c^a/k_c^b$ for $1 \le c \le m$ because $k_c^a/k_c^b$ is the utility that $a$ needs to give up in order increase $b$'s utility by one. Since $a$ wants to maximise its own utility and give $b$ a utility of $U^b([a^{t+1}, b^{t+1}], t+1)$, it divides the $m$ pies such that it gets the maximum possible share for those issues for which $k_c^a/k_c^b$ is high and gives to agent $b$ the maximum possible share for those issues for which $k_c^a/k_c^b$ is low. Thus, $a$ begins by giving $b$ the maximum possible share for the issue with the lowest $k_c^a/k_c^b$. It then does the same for the issue with the next lowest $k_c^a/k_c^b$ and repeats this process until $b$ gets a cumulative utility of $U^b([a^{t+1}, b^{t+1}], t+1)$. In order to facilitate this process of making tradeoffs, the individual elements of $k^b$ are arranged such that $k_c^a/k_c^b > k_{c+1}^a/k_{c+1}^b$. The function TRADEOFFA takes six parameters: $k^a$, $k^b$, $\delta$, UB($t$), $m$, and $t$ and uses the above described greedy method to solve the maximisation problem given in Equation 4 and return the corresponding package. If there is more than one package that solves Equation 4, then TRADEOFFA returns any one of them (because agent $a$ gets equal utility from all such packages and so does agent $b$). The function TRADEOFFB for agent $b$ is analogous to that for $a$.

On the other hand, the equilibrium strategy for the agent that receives an offer is as follows. For time period $t$, let $b$ denote the receiving agent. Then, $b$ accepts $[x^t, y^t]$ if UB($t$) $\le U^b([x^t, y^t], t)$, otherwise it rejects the offer because it can get a higher utility in the next time period. The equilibrium strategy for $a$ as receiving agent is defined analogously. Hence we get the equilibrium strategies (A($t$) and B($t$)) given in the statement of the theorem.

In this way, we reason backwards and obtain the offers for $t = 1$. The first mover makes this offer and the other agent accepts it. An agreement therefore occurs in the first time period. $\square$

**Theorem 2** *For the package deal procedure, the time taken to determine an equilibrium offer for* $t = 1$ *is* $\mathcal{O}(mn)$ *where* $m$ *is the number of issues and* $n$ *is the deadline.*

**Proof:** We know from Theorem 1 that the time to compute the equilibrium offer for $t = n$ is linear in the number of issues (see strategies A($n$) and B($n$)). Consider a time period $t < n$. During this time period, the function TRADEOFFA is used to make tradeoffs. The time complexity of TRADEOFFA (which uses the greedy approach described in the proof of Theorem 1) is $\mathcal{O}(m)$ (Martello & Toth, 1990; Cormen et al., 2003). This function needs to be repeated for every time period from the $(n-1)$th to the first. Hence the time complexity of finding an offer for the first time period is $\mathcal{O}(mn)$. $\square$

---

5. The time complexity of this approach is $\mathcal{O}(m)$ (Martello & Toth, 1990), where $m$ denotes the number of items. Note that the greedy method for the fractional knapsack problem takes $\mathcal{O}(m)$ time regardless of whether the coefficients $k_c^a$ and $k_c^b$ (for $1 \le c \le m$) in Equation 4 are positive or negative (Martello & Toth, 1990). In the present setting (as we mentioned at the beginning of Section 3) these coefficients are all positive. However, we will come across negative coefficients when we deal with interdependent issues in Section 6.

**Theorem 3** *The package deal procedure generates a Pareto optimal outcome.*

**Proof:** Recall that we consider competitive negotiations. Hence, for an individual issue $c$ (where $1 \leq c \leq m$), an increase in one agent's utility results in a decrease in that of the other. However, for the package deal procedure, an agent considers its cumulative utility from all $m$ issues. Consequently, during the process of backward reasoning, at time $t < n$, the agent that makes tradeoffs maximises its own cumulative utility without lowering that of its opponent (with respect to what the opponent would offer in the next time period). Hence the equilibrium outcome for the package deal is Pareto optimal. $\square$

**Theorem 4** *For a given first mover, the package deal procedure has a unique equilibrium outcome if the following condition is false:*

$C_1$. *There exists an $i$ and a $j$ (where $1 \leq i \leq m$ and $1 \leq j \leq m$) such that $(i \neq j)$ and $(k_i^a/k_i^b = k_j^a/k_j^b)$.*

**Proof:** Consider a time period $t < n$ and let $a$ denote the offering agent. Recall from Theorem 1 that $a$ splits the $m$ issues in the increasing order of $k_i^a/k_i^b$. Thus, for a given $i$ and $j$, if $k_i^a/k_i^b = k_j^a/k_j^b$, then agent $a$ is indifferent between which of the two issues ($i$ or $j$) it splits up first. For example, if $m = 2$, $n = 2$, $\delta = 0.5$, $k_1^a = 1$, $k_2^a = 2$, $k_1^b = 2$, and $k_2^b = 4$, then $k_i^a/k_1^b = k_2^a/k_2^b = 0.5$. If $a$ is the offering agent at $t = 1$, it can offer $(1, 0)$ for issue 1 and $(1/4, 3/4)$ for issue 2. This gives a cumulative utility of 1.5 to $a$ and 3 to $b$. Alternatively $a$ can offer $(0, 1)$ for issue 1 and $(3/4, 1/4)$ for issue 2 since this also results in the same cumulative utilities to $a$ and $b$.

On the other hand, if $k_i^a/k_i^b \neq k_j^a/k_j^b$, then $a$ splits issue $i$ first if $k_i^a/k_i^b < k_j^a/k_j^b$ and issue $j$ first if $k_i^a/k_i^b > k_j^a/k_j^b$. In other words, there is only one possible equilibrium offer that $a$ can make at any time $t < n$. Likewise there is one possible equilibrium offer that $b$ can make at any time $t < n$. Since there is a unique offer for each time period, the equilibrium outcome is unique. $\square$

Note that the uniqueness we refer to in Theorem 4 is with respect to a given first mover. If the first mover changes, then the equilibrium outcome may change, as the following example illustrates. Let $m = 2$, $n = 2$, $\delta = 0.5$, $k_1^a = 1$, $k_2^a = 2$, $k_1^b = 2$, and $k_2^b = 1$. If $a$ is the offering agent at $t = 1$, its equilibrium offer is $(1/4, 3/4)$ for the first issue and $(1, 0)$ for the second. This results in a cumulative of 2.25 to $a$ and 1.5 to $b$. In contrast, if $b$ is the offering agent at $t = 1$, its equilibrium offer is $(0, 1)$ for the first issue and $(3/4, 1/4)$ for the second. This results in a cumulative utility of 1.5 to $a$ and 2.25 to $b$. In the following discussion, we use the term unique to mean unique with respect to a given first mover.

### 3.2 The Simultaneous Procedure

For this procedure, the $m$ issues are partitioned into $\mu > 1$ disjoint subsets. For $1 \leq c \leq \mu$, let $S_c$ denote the $c$th partition where $\cup_{c=1}^{\mu} S_c = \{1, \ldots, m\}$. The issues within each subset are settled using the package deal. Negotiation for each of the $\mu$ partitions starts at $t = 1$. Thus, for $\mu = m$, all $m$ issues are settled simultaneously and independently of each other. At the other extreme, we have only one partition (i.e., $\mu = 1$) which is the package deal procedure described in Section 3.1. Since the issues in each subset (i.e., each $S_c$) are settled using the package deal, the equilibrium for each of these $\mu$ partitions is obtained from Theorem 1. Consequently, we get the following results.

First, an agreement for each issue occurs in the first round. This is because negotiation for each partition starts at $t = 1$. Also, from Theorem 1, we know that an agreement for the package deal occurs at $t = 1$. Hence, for the simultaneous procedure, an agreement for each partition (and hence each issue) occurs in the first time period.

Second, for the simultaneous procedure, the time taken to determine an equilibrium offer for $t = 1$ is $\Sigma_{c=1}^{\mu} \mathcal{O}(|S_c|n)$ where $|S_c|$ is the number of issues in the $c$th partition and $n$ is the deadline. This is explained as follows. Since the time taken to find the equilibrium offer for $t = 1$ for the package deal (i.e., for $\mu = 1$) is $\mathcal{O}(mn)$ (see Theorem 2), the time taken to compute the equilibrium offer for $t = 1$ for the $c$th partition is $\mathcal{O}(|S_c|n)$. Hence, for all $\mu$ partitions, the time complexity is $\Sigma_{c=1}^{\mu} \mathcal{O}(|S_c|n)$ which is equal to $\mathcal{O}(Mn)$, where $M$ denotes the number of issues in the largest partition.

Third, it follows from Theorem 4 that the simultaneous procedure has a unique equilibrium outcome if the following condition $C_2$ is true:

$C_2$. There is no partition $c$ (where $1 \leq c \leq \mu$) for which the condition $C_1$ is true.

Finally, as Theorem 5 shows, the simultaneous procedure may not generate a Pareto optimal outcome.

**Theorem 5** *The simultaneous procedure may not generate a Pareto optimal outcome.*

**Proof:** The package deal allows tradeoffs to be made across all the $m$ issues, while the simultaneous procedure allows tradeoffs to be made across issues within each partition but not across partitions. Hence the simultaneous procedure may not generate a Pareto optimal outcome. We show this with a counter example. Consider the case where $n = 2$, $\delta = 0.5$, $m = 3$, $\mu = 2$, $S_1 = \{1, 2\}$, $S_2 = \{3\}$, $k_1^a = 1$, $k_2^a = 2$, $k_3^a = 3$, $k_1^b = 1$, $k_2^b = 0.5$, and $k_3^b = 0.25$. Let $a$ denote the first mover. From Theorem 1, we know that in the equilibrium for partition $S_1$, agent $a$ gets a share of 0.25 for issue 1 and 1 for issue 2, and $b$ gets a share of 0.75 for issue 1 and nothing for issue 2. For partition $S_2$, each agent gets a share of $1/2$. Thus, $a$'s cumulative utility from all three issues is 3.75 and that of $b$ is 0.875.

Now consider the case where all three issues are discussed using the package deal. Here, $\mu = 1$ and all other parameters remain the same. In the equilibrium outcome for this procedure, $a$ gets a cumulative utility of 5.125 and $b$ gets 0.875. This means that the procedure with $\mu = 2$ does not generate a Pareto optimal outcome. $\square$

### 3.3 The Sequential Procedure

For this procedure, the $m$ issues are partitioned into $\mu > 1$ disjoint subsets. For $1 \leq c \leq \mu$, let $S_c$ denote the $c$th partition where $\cup_{c=1}^{\mu} S_c = \{1, \ldots, m\}$. The $\mu$ partitions are negotiated sequentially, one after another. The issues within a subset are settled using the package deal. Negotiation for the first partition starts at time $t = 1$. If negotiation for the $c$th (for $1 \leq c \leq \mu$) partition ends at $t_c$, then negotiation for the $(c + 1)$th partition starts at time $t_c + 1$. Each player gets its share for all the issues in a partition as soon as the partition is settled. Thus, for $\mu = m$, all $m$ issues are settled in sequence. At the other extreme, we have only one partition (i.e., $\mu = 1$) which is the package deal procedure described in Section 3.1. Since the issues in each subset (i.e., each $S_c$) are settled using the package deal, the equilibrium for each of these $\mu$ subsets is obtained from Theorem 1 by substituting the appropriate negotiation start times for each partition.

**Theorem 6** *For the sequential procedure, the equilibrium time of agreement for the $c$th partition (for $1 \leq c \leq \mu$) is $T_c = c$.*

**Proof:** From Theorem 1, we know that an agreement for the package deal occurs in the first time period. Hence, negotiation for each partition ends in the same time period in which it starts (i.e., negotiation for the $c$th partition starts at $t = c$ and results in an agreement in the same time period). The time taken to settle all the $m$ issues is therefore $\mu$. $\square$

Note that the time complexity of the sequential procedure (i.e., the time to compute equilibrium offers) is the same as that for the simultaneous procedure. Also, like the simultaneous procedure, the equilibrium outcome for the sequential procedure may not be Pareto optimal. Finally, the condition for the equilibrium outcome for the sequential procedure to be unique is the same as that for the simultaneous procedure.

### 3.4 The Optimal Procedure

Having obtained the equilibrium outcomes for the three multi-issue procedures, we now compare them in terms of the utilities they generate for each player. Then the procedure that gives a player the maximum utility is its optimal one.

Note that, for the sequential procedure, the equilibrium outcome strongly depends on the order in which the partitions are settled. This ordering is called the negotiation *agenda*. There are two ways of defining the agenda (Fershtman, 1990): *exogenously* or *endogenously*. If the agenda is determined before the actual negotiation over the issues begins, then it is said to be exogenous. On the other hand, for the endogenous agenda, the agents decide what issue they will settle next during the process of negotiation. The agenda that gives an agent the maximum utility between all possible agendas is its optimal one (Fatima et al., 2004). Our objective here is not to determine the optimal agenda, but to consider a given agenda and compare the equilibrium outcome for the sequential procedure for that agenda with the outcomes for the simultaneous and the package deal procedures, in order to find the optimal procedure. The following theorem characterises this procedure.

**Theorem 7** *Irrespective of how the $m$ issues are split into $\mu > 1$ partitions, the package deal is optimal for both parties.*

**Proof:** In order to compare an agent's utility from different procedures, it is important to take into account who initiates negotiation. For the package deal, the first mover makes an offer on all the issues. Hence we compare an agent's utilities for the three procedures, given the agent that will be the first mover for all the three procedures for all the issues.

We first show that the outcome for the package deal is no worse than that for the simultaneous procedure. Consider the simultaneous procedure for any $\mu > 1$. For this procedure, for $t \leq n$, the offering agent makes tradeoffs across the issues in each partition independently of the other partitions. Now consider the package deal procedure (i.e., with $\mu = 1$ partitions). For this procedure, the offering agent makes tradeoffs across all $m$ issues. Since the difference between the procedure with $\mu = 1$ and the one with $\mu > 1$ is that the former makes tradeoffs across all $m$ issues while the latter does not, each agent's utility from the former procedure is no worse than its utility from the latter.

We now show that for a given $\mu$ (where $\mu > 1$), for each agent, the outcome for the simultaneous procedure is better than that for the sequential one (irrespective of the agenda for the sequential procedure). We do this by considering each of the $\mu$ partitions.

| | Package deal | Simultaneous | Sequential |
|---|---|---|---|
| Time of agreement ($t_c$) | For the $c$th issue $t_c = 1$ for $1 \leq c \leq m$ | For the $c$th issue $t_c = 1$ for $1 \leq c \leq m$ | For the $c$th partition $t_c = c$ for $1 \leq c \leq \mu$ |
| Time to compute equilibrium | $\mathcal{O}(mn)$ | $\mathcal{O}(Mn)$ | $\mathcal{O}(Mn)$ |
| Pareto optimal? | Yes | No | No |
| Unique equilibrium? | If $\neg C_1$ | If $C_2$ | If $C_2$ |

Table 2: A comparison of the outcomes for the three multi-issue procedures for the complete information setting ($CI$).

- Partition $c = 1$. Since negotiation for the first partition starts at $t = 1$ for both the simultaneous and the sequential procedures, the outcome for this partition is the same for $\mu = 1$ and $\mu > 1$. Hence, for the first partition, an agent gets equal utility from the two procedures.

- Partition $c > 1$. Let agent $a$ denote the first mover for partition $c$ (for $2 \leq c \leq \mu$) for both simultaneous and sequential procedures. Also, let $U_{sim}^a$ and $U_{seq}^a$ denote $a$'s cumulative utility for this partition from the equilibrium outcome for the simultaneous and the sequential procedures respectively. Likewise, let $U_{sim}^b$ and $U_{seq}^b$ denote $b$'s cumulative utility for this partition from the equilibrium outcome for the simultaneous and the sequential procedures respectively.

  Now for the simultaneous procedure, negotiation for each partition starts in the first time period. An agreement for each partition also occurs in the first time period. On the other hand, for the sequential procedure, negotiation for the $c$th partition starts in the $c$th time period and results in an agreement in the same time period (see Theorem 6). Since each pie shrinks with time, agent $a$'s cumulative utility $U_{sim}^a$ is greater than $U_{seq}^a$, and agent $b$'s cumulative utility $U_{sim}^b$ is greater than $U_{seq}^b$.

Thus, the simultaneous procedure is better than the sequential one for both agents. Furthermore (as shown above), the outcome for the package deal is no worse than that for the simultaneous procedure for both agents. Therefore, for each agent, the package deal is the optimal procedure. □

These results are summarised in Table 2. For the above analysis, the negotiation parameters $n$, $\delta_c$, $k_c^a$, and $k_c^b$ (for $1 \leq c \leq m$) were common knowledge to the agents. However, this is unlikely to be the case for most encounters. Therefore we now extend this analysis to incomplete information scenarios with uncertainty about utility functions[6]. In Section 4, we focus on the symmetric information setting where each agent is uncertain about the other's utility function. Then, in Section 5, we examine the asymmetric information setting where one of the two agents is uncertain about the other's utility function, but the other agent knows the utility function of both agents.

---

6. There are two other sources of uncertainty: uncertainty about the negotiation deadline and uncertainty about discount factors. Future work will deal with uncertainty about discount factors. However, for independent issues, we analysed the case with symmetric uncertainty about deadlines in (Fatima, Wooldridge, & Jennings, 2006). The extension of this work to the case of interdependent issues is another direction for future work.

## 4. Multi-Issue Negotiation with Symmetric Uncertainty about the Opponent's Utility

In this symmetric information setting, each agent is uncertain about its opponent's utility function: for $1 \leq c \leq m$, agent $a$ ($b$) is uncertain about $k_c^b$ ($k_c^a$). Specifically, let $K$ denote a vector of $r$ vectors where each vector $K_i \in \mathbb{R}_+^m$ (for $1 \leq i \leq r$) consists of $m$ constant positive real numbers. These $r$ vectors are the possible values for $k^a \in \mathbb{R}_+^m$ and $k^b \in \mathbb{R}_+^m$. In other words, there are $r$ types[7] for agent $a$ and $r$ types for agent $b$. Let $P^a : \mathbb{N}^+ \to \mathbb{R}_1$ denote the discrete probability distribution function for $k^a$ and $P^b : \mathbb{N}^+ \to \mathbb{R}_1$ that for $k^b$. The domain for these two functions is $[1..r]$. In other words, for $1 \leq i \leq r$, $P^a(i)$ ($P^b(i)$) is the probability that agent $a$ ($b$) is of type $i$. For $1 \leq c \leq m$, let $K_{ic}$ denote the $c$th element of vector $K_i$.

In this setting, the vector $K$ and the functions $P^a$ and $P^b$ are common knowledge to the negotiators. Also, each agent knows its own type, but not that of its opponent. In addition, each agent knows $r$, $\delta$, $n$, and $m$.

Since there are $r$ types for agent $a$ and $r$ types for agent $b$, we define $r$ different cumulative utility functions for each of the two agents. If agent $a$ ($b$) is of type $i$ (for $1 \leq i \leq r$) then its utility $U_i^a : \mathbb{R}_1^m \times \mathbb{R}_1^m \times \mathbb{N}^+ \to \mathbb{R}$ ($U_i^b : \mathbb{R}_1^m \times \mathbb{R}_1^m \times \mathbb{N}^+ \to \mathbb{R}$) from the division specified by the package $[x^t, y^t]$ at time $t$ is:

$$U_i^a([x^t, y^t], t) = \begin{cases} \Sigma_{c=1}^m K_{ic} u_c^a(x_c^t, t) & \text{if } t \leq n \\ 0 & \text{otherwise} \end{cases} \qquad (5)$$

$$U_i^b([x^t, y^t], t) = \begin{cases} \Sigma_{c=1}^m K_{ic} u_c^b(y_c^t, t) & \text{if } t \leq n \\ 0 & \text{otherwise} \end{cases} \qquad (6)$$

Note that, as before, the issues are perfect substitutes. For this setting, we determine the equilibrium outcomes for each of the three multi-issue procedures and then compare them.

### 4.1 The Package Deal Procedure

We know from Theorem 1 that the equilibrium outcome for the complete information setting depends on $k_c^a$ and $k_c^b$ (for $1 \leq c \leq m$). However, in this setting, there is uncertainty about $k_c^a$ and $k_c^b$. Hence we use the standard expected utility theory (Neumann & Morgenstern, 1947; Fishburn, 1988; Harsanyi & Selten, 1972) to find an agent's optimal strategy. Before doing so, however, we first introduce some notation.

For $1 \leq i \leq r$, we let $\text{A}(i, t)$ denote the equilibrium strategy for an agent $a$ of type $i$ for the time period $t$. Analogously, $\text{B}(i, t)$ denotes the equilibrium strategy for an agent $b$ of type $i$ for the time period $t$. Note that for $1 \leq i \leq r$, if $[a^t, b^t]$ is the package offered at time $t$ in equilibrium, then $a^t + b^t = \delta^{t-1}$ (i.e., for each pie, the sum of the shares of the two agents is equal to the size of the pie at time $t$). Also, for $1 \leq i \leq r$, we let $\text{A}(i, j, t)$ denote the equilibrium strategy for an agent $a$ of type $i$ for the time period $t$, assuming that $b$ is of type $j$. Analogously, $\text{B}(i, j, t)$ denotes the equilibrium strategy for an agent $b$ of type $i$ for the time period $t$, assuming that $a$ is of type $j$.

Also, let $\text{EUA}(i, t)$ denote the cumulative utility that an agent $a$ of type $i$ expects to get from $b$'s equilibrium offer at time $t$ (i.e., $a$ is the receiving agent and $b$ the offering agent at $t$). Likewise, $\text{EUB}(i, t)$ denotes the cumulative utility that an agent $b$ of type $i$ expects to get from $a$'s equilibrium offer at time $t$ (i.e., $b$ is the receiving agent and $a$ the offering agent at $t$). We let $\text{EUA}(i, j, t)$ denote agent $a$'s expected cumulative utility from its own equilibrium offer at time $t$ if $a$ is of type $i$, assuming that $b$ is of type $j$. Note that this is $a$'s utility when it is the offering agent at $t$. And let $\textsc{eub}(i,j,t)$ denote agent $b$'s expected cumulative utility from its own equilibrium offer at time $t$ if $b$ is of type $i$ and assuming that $a$ is of type $j$. Note that this is $b$'s utility when it is the offering agent at $t$.

---

7. An agent's type indicates which of the $r$ vectors it corresponds to.

Recall that in this setting, each agent only knows its own type, but not that of its opponent. Since there are $r$ possible types, there are $r$ possible offers an agent can make at any time period (one offer corresponding to each of the opponent's types). Among these $r$ offers, the one that gives an agent the maximum expected cumulative utility is its optimal offer. If the $c$th offer ($1 \leq c \leq r$) gives an agent the maximum expected cumulative utility, then we say that the *optimal choice* for the agent is $c$. For time period $t$, we let $\textsc{opta}(i,t)$ ($\textsc{optb}(i,t)$) denote the optimal choice for agent $a$ ($b$) of type $i$.

At $t = n$, the offering agent gets everything and the opponent gets zero utility. Thus, for $t = n$, we have the following:

$$\textsc{eua}(i,n) = 0 \qquad \text{for } 1 \leq i \leq r \tag{7}$$

$$\textsc{eub}(i,n) = 0 \qquad \text{for } 1 \leq i \leq r \tag{8}$$

$$\textsc{eua}(i,j,n) = \sum_{c=1}^{m} K_{ic}\delta_c^{t-1} \qquad \text{for } 1 \leq i \leq r \text{ and } 1 \leq j \leq r \tag{9}$$

$$\textsc{eub}(i,j,n) = \sum_{c=1}^{m} K_{ic}\delta_c^{t-1} \qquad \text{for } 1 \leq i \leq r \text{ and } 1 \leq j \leq r \tag{10}$$

Note that for $t = n$, $\textsc{eua}(i,j,n)$ and $\textsc{eub}(i,j,n)$ do not depend on $j$ because in the last time period, the offering agent gets 100 percent of all the $m$ pies. For all preceding time periods $t < n$, we have the following:

$$\textsc{eua}(i,t) = \textsc{eua}(i,\theta,t+1) \qquad \text{for } 1 \leq i \leq r \text{ where } \theta = \textsc{opta}(i,t+1) \tag{11}$$

$$\textsc{eub}(i,t) = \textsc{eub}(i,\lambda,t+1) \qquad \text{for } 1 \leq i \leq r \text{ where } \lambda = \textsc{optb}(i,t+1) \tag{12}$$

$$\textsc{eua}(i,j,t) = \sum_{e=1}^{r} F^a(i,j,e,t) \times P^b(e) \qquad \text{for } 1 \leq i \leq r \text{ and } 1 \leq j \leq r \tag{13}$$

$$\textsc{eub}(i,j,t) = \sum_{e=1}^{r} F^b(i,j,e,t) \times P^a(e) \qquad \text{for } 1 \leq i \leq r \text{ and } 1 \leq j \leq r \tag{14}$$

The function $F^a$ takes four parameters: $i$, $j$, $e$, and $t$, and returns the utility that an agent $a$ of type $i$ gets from offering the equilibrium package for time $t$, assuming that agent $b$ is of type $j$ where in fact it is of type $e$. Obviously, agent $b$ accepts $a$'s offer at $t$ if $U_e^b(\textsc{a}(i,j,t),t) \geq \textsc{eub}(e,\gamma,t+1)$ where $\gamma = \textsc{optb}(e,t+1)$. Otherwise, agent $b$ rejects $a$'s offer and negotiation proceeds to the next round in which case $a$'s expected utility is $\textsc{eua}(i,t+1)$. Hence, $F^a$ is defined as follows:

$$F^a(i,j,e,t) = \begin{cases} U_i^a(\textsc{a}(i,j,t),t) & \text{if } U_e^b(\textsc{a}(i,j,t),t) \geq \textsc{eub}(e,\gamma,t+1) \text{ where } \gamma = \textsc{optb}(e,t+1) \\ \textsc{eua}(i,t+1) & \text{otherwise} \end{cases}$$

where the strategy $\textsc{a}(i,j,t)$ for $t = n$ is defined as follows:

$$\textsc{a}(i,j,n) = \begin{cases} \text{OFFER } [\delta^{n-1}, \mathbf{0}] & \text{if } a\text{'s turn} \\ \text{ACCEPT} & \text{otherwise} \end{cases}$$

and for all preceding time periods $t < n$ it is defined as:

$$\text{A}(i,j,t) = \begin{cases} \text{OFFER TRADEOFFA1}(K, \delta, \text{EUB}(j,t), i, j, m, t, P^a, P^b) & \text{if } a\text{'s turn} \\ \text{if } U_i^a([x^t, y^t], t) \geq \text{EUA}(i,t) \text{ ACCEPT else REJECT} & \text{otherwise} \end{cases}$$

where $[x^t, y^t]$ denotes the offer made at $t$ and the function[8] TRADEOFFA1 is defined as follows. Like TRADEOFFA, the function TRADEOFFA1 solves the following maximisation problem:

$$\begin{aligned} \text{maximise} \quad & \Sigma_{c=1}^{m} K_{ic} a_c^t \\ \text{such that} \quad & \Sigma_{c=1}^{m} (\delta_c^{t-1} - a_c^t) K_{jc} = \text{EUB}(j,t) \\ & 0 \leq a_c^t \leq 1 \qquad \text{for } 1 \leq c \leq m \end{aligned} \tag{15}$$

where $i$ denotes $a$'s type and $j$ that of $b$. However, the difference between TRADEOFFA1 and TRADEOFFA arises when there is more than one package that maximises $a$'s cumulative utility (i.e., $\Sigma_{c=1}^{m} K_{ic} a_c^t$) while giving $b$ a cumulative utility of $\text{EUB}(j,t)$. If there is more than one such package, then in Theorem 1, it does not matter which of these packages $a$ offers to $b$ (because both agents have complete information). Hence, TRADEOFFA can return any one such package. However, in the present setting, there is uncertainty. Therefore, if there is more than one package that maximises $a$'s cumulative utility while giving $b$ a cumulative utility of $\text{EUB}(j,t)$, then TRADEOFFA1 returns the package that maximises $a$'s expected cumulative utility. For instance, let $[a^t, b^t]$ be one such package that maximises $a$'s cumulative utility. Then $a$'s expected cumulative utility from $[a^t, b^t]$ (i.e., $\text{EUA}(i,j,t)$) is as given in Equation 13 where:

$$F^a(i,j,e,t) = \begin{cases} U_i^a([a^t, b^t], t) & \text{if } U_e^b([a^t, b^t], t) \geq \text{EUB}(e, \gamma, t+1) \text{ where } \gamma = \text{OPTB}(e, t+1) \\ \text{EUA}(i, t+1) & \text{otherwise} \end{cases}$$

Obviously, if there is more than one package that maximises $a$'s expected cumulative utility and gives $b$ a utility of $\text{EUB}(j,t)$ then TRADEOFFA1 returns any one such package.

We now turn to agent $b$. For this agent, $F^b$, B$(i,j,t)$, and TRADEOFFB1 are defined analogously as follows:

$$F^b(i,j,e,t) = \begin{cases} U_i^b(\text{B}(i,j,t), t) & \text{if } U_e^a(\text{B}(i,j,t), t) \geq \text{EUA}(e, \alpha, t+1) \text{ where } \alpha = \text{OPTA}(e, t+1) \\ \text{EUB}(i, t+1) & \text{otherwise} \end{cases}$$

where the strategy B$(i,j,t)$ for $t = n$ is defined as follows:

$$\text{B}(i,j,n) = \begin{cases} \text{OFFER } [\mathbf{0}, \delta^{n-1}] & \text{if } b\text{'s turn} \\ \text{ACCEPT} & \text{otherwise} \end{cases}$$

and for all preceding time periods $t < n$ it is defined as:

$$\text{B}(i,j,t) = \begin{cases} \text{OFFER TRADEOFFB1}(K, \delta, \text{EUA}(j,t), i, j, m, t, P^a, P^b) & \text{if } b\text{'s turn} \\ \text{if } U_i^b([x^t, y^t], t) \geq \text{EUB}(i,t) \text{ ACCEPT else REJECT} & \text{otherwise} \end{cases}$$

---

8. A method for making tradeoffs has been proposed by Faratin, Sierra, and Jennings (2002) for an incomplete information setting, but this method differs from ours. Also, Faratin et al. only present a method for making tradeoffs, but they do not show that the resulting offer is in equilibrium. In contrast, our method shows that the resulting offer is in equilibrium.

Thus, the optimal choice for agent $a$ (i.e., $\text{OPTA}(i, t)$) and that for agent $b$ (i.e., $\text{OPTB}(i, t)$) are defined as follows:

$$\text{OPTA}(i, t) = \arg \max_{j=1}^{r} \text{EUA}(i, j, t) \quad \text{for } 1 \leq i \leq r \quad (16)$$

$$\text{OPTB}(i, t) = \arg \max_{j=1}^{r} \text{EUB}(i, j, t) \quad \text{for } 1 \leq i \leq r \quad (17)$$

Note that the offering agent's optimal choice for $t = n$ does not depend on its opponent's type since the offering agent gets all the pies.

We compute the optimal choice for the first time period by reasoning backwards from $t = n$. At $t = 1$, if an agent $a$ of type $i$ is the offering agent, then it offers the package that corresponds to agent $b$ being of type $\text{OPTA}(i, 1)$. Likewise, if an agent $b$ of type $i$ is the offering agent, then it offers the package that corresponds to agent $a$ being of type $\text{OPTB}(i, 1)$.

However, since $\text{OPTA}(i, 1)$ and $\text{OPTB}(i, 1)$ are obtained in the absence of complete information, an agreement may or may not take place in the first time period. If an agreement does not occur at $t = 1$, then the agents need to update their beliefs as follows. Let $T_t^a \subseteq \{1, 2, \ldots, r\}$ denote the set of possible types for agent $a$ at time $t$. For $t = 1$, we have $T_1^a = \{1, 2, \ldots, r\}$ and $T_1^b = \{1, 2, \ldots, r\}$. Assume that an agent $a$ of type $i$ makes an offer at $t = 1$. If the offer that $a$ makes gets rejected, then it means that $b$ is not of type $\text{OPTA}(i, 1)$ and so $a$ updates its beliefs about $b$ using Bayes' rule. Now, on the basis of $a$'s offer at $t = 1$ (say $[x^1, y^1]$), agent $b$ can infer the possible types for agent $a$. Thus, agent $b$ too updates its beliefs using Bayes' rule. The belief update rules for time $t$ are as defined below.

> UPDATE BELIEFS: Agent $a$ puts all the weight of the posterior distribution of $b$'s type over $T_t^b - \{\text{OPTB}(i, t)\}$ using Bayes' rule. Agent $b$ puts all the weight of the posterior distribution of $a$'s type over $\mathcal{K}$ using Bayes' rule where $\mathcal{K} \subseteq \{1, 2, \ldots, r\}$ is the set of possible types for $a$ that can offer $[x^t, y^t]$ in equilibrium.

The belief update rule for the case where $b$ offers at $t = 1$ is analogous to the above case where $a$ offers at $t = 1$.

Thus if the offer at $t = 1$ gets rejected, then negotiation goes to the next round. At $t = 2$, the offering agent (say an agent $a$ of type $i$) finds $\text{OPTA}(i, 2)$ with its updated beliefs. This process of updating beliefs and making offers continues until an agreement is reached.

In Section 3, we used the concept of Nash equilibrium because the agents had complete information. However, in the current setting, each agent is uncertain about its opponent's type and so an agent's optimal strategy depends on its beliefs about its opponent. Hence we use the concept of *sequential equilibrium* (Kreps & Wilson, 1982; van Damme, 1983) for this setting. Sequential equilibrium is defined in terms of two elements: a *strategy profile* and a *system of beliefs*. The strategy profile comprises of a pair of strategies, one for each agent. The belief system has the following properties. Each agent has a belief about its opponent's type. In each time period, an agent's strategy is optimal given its current beliefs (during the time period) and the opponent's possible strategies. For each time period, each agent's beliefs (about its opponent) are consistent with the offers it received. Using this concept of sequential equilibrium, the following theorem characterises the equilibrium for the package deal procedure.

**Theorem 8** *For the package deal procedure, the following strategies form a sequential equilibrium. The equilibrium strategies for $t = n$ are:*

$$\text{A}(i, n) = \begin{cases} \text{OFFER } [\delta^{n-1}, \boldsymbol{0}] & \text{IF } a\text{'s TURN} \\ \text{ACCEPT} & \text{IF } b\text{'s TURN} \end{cases}$$

$$\text{B}(i,n) = \left\{ \begin{array}{ll} \textit{OFFER } [\boldsymbol{0}, \delta^{n-1}] & \textit{IF b's TURN} \\ \textit{ACCEPT} & \textit{IF a's TURN} \end{array} \right.$$

*for $1 \leq i \leq r$. For all preceding time periods $t < n$, if $[x^t, y^t]$ denotes the offer made at time $t$, then the equilibrium strategies are defined as follows:*

$$\text{A}(i,t) = \left\{ \begin{array}{ll} \textit{OFFER } \text{TRADEOFFA1}(K, \delta, \text{EUB}(\psi, t), i, \psi, m, t, P^a, P^b) & \textit{IF a's TURN} \\ \textit{If offer gets rejected UPDATE BELIEFS} & \\ \textit{RECEIVE OFFER and UPDATE BELIEFS} & \textit{IF b's TURN} \\ \textit{If } (U_i^a([x^t, y^t], t) \geq \text{EUA}(i,t)) \textit{ ACCEPT else REJECT} & \end{array} \right.$$

$$\text{B}(i,t) = \left\{ \begin{array}{ll} \textit{OFFER } \text{TRADEOFFB1}(K, \delta, \text{EUA}(\phi, t), i, \phi, m, t, P^a, P^b) & \textit{IF b's TURN} \\ \textit{If offer gets rejected UPDATE BELIEFS} & \\ \textit{RECEIVE OFFER and UPDATE BELIEFS} & \textit{IF a's TURN} \\ \textit{If } (U_i^b(x^t, y^t], t) \geq \text{EUB}(i,t)) \textit{ ACCEPT else REJECT} & \end{array} \right.$$

*for $1 \leq i \leq r$. Here, $\psi = \text{OPTA}(i,t)$ and $\phi = \text{OPTB}(i,t)$. The earliest possible time of agreement is $t = 1$ and the latest possible time of agreement is $t = min(2r - 1, n)$.*

**Proof:** At time $t = n$, the offering agent takes all the pies and leaves nothing for its opponent. The opponent accepts this and we get $\text{A}(i,n)$ and $\text{B}(i,n)$. Now consider a time period $t < n$. Recall that during negotiation for the complete information setting (see Section 3.1), at time $t < n$, the offering agent proposes a package that gives its opponent a cumulative utility equal to what the opponent would get from its own equilibrium offer for the next time period. However, for the current incomplete information setting, an agent knows its own type but not that of its opponent. Hence, for this scenario, at time $t < n$, the offering agent (say $a$) proposes a package that gives $b$ an expected cumulative utility equal to what $b$ would get from its own equilibrium offer for the next time period (i.e., $\text{EUB}(\psi, t)$). This package is determined by the TRADEOFFA1 function. Likewise, if $b$ is the offering agent at time $t$, then it makes tradeoffs using TRADEOFFB1 and offers $a$ an expected cumulative utility $\text{EUA}(\phi, t)$.

We obtain the equilibrium offer for $t = n - 1$ and then reason backwards until we obtain the equilibrium offer for $t = 1$. However, since these offers are computed in the absence of complete information (i.e., on the basis of expected utilities), an agreement may or may not take place at $t = 1$. If an agreement does not take place at $t = 1$, then negotiation proceeds as follows. Consider a time period $t$ such that $1 \leq t < n$. Let $[x^t, y^t]$ denote the offer made at time $t$. The agent that receives the offer (say agent $a$) updates its beliefs using Bayes' rule: put all the weight of the posterior distribution of $b$'s type over $\mathcal{K}$ where $\mathcal{K} \subseteq \{1, 2, \ldots, r\}$ is the set of possible types for $b$ that can offer $[x^t, y^t]$ in equilibrium. If the proposed offer ($[x^t, y^t]$) gets rejected, then the offering agent (say agent $b$ of type $i$) updates its beliefs using Bayes' rule: put all the weight of the posterior distribution of $a$'s type over $T_t^a - \{\text{OPTB}(i,t)\}$. The belief update rule for the case where agent $a$ offers at time $t$ are analogous to the above rule. These belief update rules when incorporated in the agents' strategies give $\text{A}(i,t)$ and $\text{B}(i,t)$ as shown in the statement of the theorem.

We now show that the beliefs specified above are consistent. During any time period $t < n$, let the strategy profile ($\text{A}(i,t)$, $\text{B}(i,t)$) assign probability $1 - \epsilon$ to the above specified posterior beliefs and probability $\epsilon$ to the rest of the support for the opponent's type. As $\epsilon \rightarrow 0$, the above strategy pair converges to ($\text{A}$, $\text{B}$). Also, the beliefs generated by the strategy pair converge to the beliefs described above. Given these beliefs, the strategies $\text{A}$ and $\text{B}$ are sequentially rational.

The earliest possible time of agreement is $t = 1$. We show this with the following example. Let $n = 2$, $m = 2$, $r = 2$, $\delta = 1/2$, and $K = [1, 2; 5, 1]$. Let agent $a$ be the offering agent at time $t = 1$. Assume that $a$ is of type 1 (i.e., $k^a = [1, 2]$). Let $P^b(1) = 0.1$ and $P^b(2) = 0.9$. Since $r = 2$, agent $a$ can play two possible strategies at time $t = 1$: one that corresponds to the case where $b$ is of type 1 and the other that corresponds to the case where $b$ is of type 2. For the former case, $a$'s equilibrium offer at $t = 1$ is $[0, 1]$ for the first issue and $[\frac{3}{4}, \frac{1}{4}]$ for the second one. Hence $\text{EUA}(1, 1, 1) = 1.5$. For the latter case, $a$'s equilibrium offer at $t = 1$ is $[\frac{2}{5}, \frac{3}{5}]$ for the first issue and $[1, 0]$ for the second issue. Hence $\text{EUA}(1, 2, 1) = 2.16$. Since $\text{EUA}(1, 2, 1) > \text{EUA}(1, 1, 1)$, $\text{OPTA}(1, 1) = 2$ and $a$ plays the latter strategy. Now if $b$ is in fact of type 2, then it accepts $a$'s offer at $t = 1$. But if $b$ is in fact of type 1, it rejects $a$'s offer at $t = 1$ since it can get a higher utility at $t = 2$. An agreement therefore occurs at $t = 2$. Thus, the earliest possible time of agreement is $t = 1$.

Now consider the case where an $a$ of type $i$ offers at $t = 1$ but an agreement does not occur at this time. When $a$'s offer gets rejected, it knows that $b$ is not of type $\text{OPTA}(i, 1)$. Thus the number of possible types for $b$ is now reduced to $r - 1$. This happens every time $a$ makes an offer (i.e., every alternate time period) but it gets rejected. When negotiation reaches time period $t = 2r - 1$, there is only one possible type for $b$. Likewise, there is only one possible type for agent $a$. An agreement therefore takes place at $t = 2r - 1$. However, if $n < 2r - 1$ then an agreement occurs at $t = n$ (see $\text{A}(i, n)$ and $\text{B}(i, n)$). In other words, if an agreement does not occur at $t = 1$, then it occurs at the latest by $t = min(2r - 1, n)$. $\square$

As we mentioned earlier, if there is more than one package that solves Equation 15, then $\text{TRADEOFFA1}$ returns the one that maximises $a$'s expected cumulative utility. Let $\text{PA}_t^{ij}$ (where $i$ denotes $a$'s type and $j$ that of $b$) denote the set of all possible packages that $\text{TRADEOFFA1}$ can return at time $t$. The set $\text{PB}_t^{ij}$ for agent $b$ is defined analogously.

**Theorem 9** *For a given first mover, the package deal procedure has a unique equilibrium outcome if the condition $C_3$ is false or $C_4$ is true.*

$C_3$. *There exists an $i$, $j$, $c$, and $d$, such that $(c \neq d)$ and $(i \neq j)$ and $(K_{ic}/K_{jc} = K_{id}/K_{jd})$ where $1 \leq i \leq r$, $1 \leq j \leq r$, $1 \leq c \leq m$, and $1 \leq d \leq m$.*

$C_4$. $|\text{PA}_t^{ij}| = 1$ *and* $|\text{PB}_t^{ij}| = 1$ *where* $1 \leq i \leq r$, $1 \leq j \leq r$, $i \neq j$, *and* $1 \leq t \leq n$.

**Proof:** Let $i$ denote agent $a$'s type and $j$ denote $b$'s type where $i \neq j$, $1 \leq i \leq r$, and $1 \leq k \leq r$. Note that if $a$ and $b$ are of the same type, they have similar preferences for different issues. So $i \neq j$ because the agents gain from making tradeoffs when they are of different types. The rest of the proof for the condition $C_3$ follows from Theorem 4. Consider $C_4$. If $C_3$ is true, then we know that, at time $t$, $\text{TRADEOFFA1}$ returns that package that solves Equation 15 and maximises $a$'s expected cumulative utility. Hence if $\text{PA}_t^{ij}$ contains a single element, then there is only one possible package that $\text{TRADEOFFA1}$ can return. Likewise, if $\text{PB}_t^{ij}$ contains a single element, then there is only one possible package that $\text{TRADEOFFB1}$ can return. If there is only one possible offer for each time period $1 \leq t \leq n$, then the equilibrium outcome is unique. $\square$

In order to determine the time complexity of the package deal, we first find the complexity of the $\text{TRADEOFFA1}$ function. As we mentioned before, $\text{TRADEOFFA1}$ differs from $\text{TRADEOFFA}$ when there is more than one package that solves the maximisation problem of Equation 15. We know from Theorem 9 that there is more than one such package if the condition $C_3$ is true. We also know from Theorem 1 that using the greedy approach, TRADEOFFA considers the $m$ issues in the increasing order of $K_{ic}/K_{jc}$ where $i$ denotes $a$'s type and $j$ denotes $b$'s type. Let $\mathcal{S}_p^{ij} \subseteq S$ denote a set of issues (where $0 \leq D^{ij} < m$, $1 \leq p \leq D^{ij}$, $i$ denotes $a$'s type, and $j$ denotes $b$'s type) such that:

$$|\mathcal{S}_p^{ij}| > 1 \quad \text{for } 1 \leq p \leq D^{ij}$$

and:

$$\forall_{c,d \in \mathcal{S}_p^{ij}} \frac{K_{ic}}{K_{jc}} = \frac{K_{id}}{K_{jd}}$$

In other words, $\mathcal{S}_p^{ij}$ is a set of issues such that if $c$ and $d$ belong to $\mathcal{S}_p^{ij}$ then $K_{ic}/K_{jc} = K_{id}/K_{jd}$, and $D^{ij}$ is the number of sets that satisfy this condition. So if $D^{ij} = 0$ then it means that there is only one package that solves Equation 15. But if $D^{ij} > 0$ then there is more than one package that solves Equation 15 and from among these TRADEOFFA1 must find the one that maximises $a$'s expected cumulative utility. For example if the set of issues is $S = \{1, 2, 3, 4\}$, $r = 2$, $K_1 = \{5, 6, 7, 8\}$, and $K_2 = \{9, 6, 7, 8\}$, then $D^{12} = 1$, $\mathcal{S}_1^{12} = \{2, 3, 4\}$, and $|\mathcal{S}_1^{12}| = 3$. So while making tradeoffs, $a$ can consider the issues in $\mathcal{S}_1^{12}$ in any order because for all the three issues it needs to give up the same amount of utility in order to increase $b$'s utility by 1. The three issues in $\mathcal{S}_1^{12}$ can be ordered in 3! different ways resulting in 3! different packages. From among these 3! different packages, TRADEOFFA1 must find the one that maximises $a$'s expected cumulative utility. In general, for $D^{ij} > 1$, let $\pi^{ij}$ denote the number[9] of possible packages TRADEOFFA1 needs to consider where $\pi^{ij}$ is:

$$\pi^{ij} = \prod_{p=1}^{D^{ij}} |\mathcal{S}_p^{ij}|!$$

In other words, if $a$'s type is $i$ and $b$'s type is $j$, then there are $\pi^{ij}$ packages that solve Equation 15 and from among these TRADEOFFA1 must find the one that maximises $a$'s expected cumulative utility. So if $D^{ij} = 0$, then $\pi^{ij} = 1$. Let $\hat{\pi}$ be defined as:

$$\hat{\pi} = \max_{1 \leq i \leq r, 1 \leq j \leq r, i \neq j} \pi^{ij} \tag{18}$$

In other words, $\hat{\pi}$ is the maximum number of packages that TRADEOFFA1 will have to search to find the one that maximises $a$'s expected cumulative utility (considering all possible types of $a$ and all possible types of $b$). Note that, as before, $a$ and $b$ are of different types (i.e., $i \neq j$ in Equation 18) because the agents gain from making tradeoffs when they are of different types. The time complexity of TRADEOFFA1 depends on $\hat{\pi}$.

**Theorem 10** *The time complexity of* TRADEOFFA1 *is* $\mathcal{O}(m\hat{\pi})$.

**Proof:** We know from Theorem 2 that the time complexity of finding any one package that solves Equation 15 is $\mathcal{O}(m)$. However, if there is more than one package that solves Equation 15 then TRADEOFFA1 returns the one that maximises $a$'s expected cumulative utility. The time to compute $a$'s expected cumulative utility from any one such package is $\mathcal{O}(m)$. The maximum number of such packages for which $a$ needs to find its expected cumulative utility is $\hat{\pi}$. Thus the time complexity of TRADEOFFA1 is $\mathcal{O}(m\hat{\pi})$. $\square$

---

9. Note that $\pi^{ij}$ is defined in terms of the factorial of $|\mathcal{S}_p^{ij}|$, but $|\mathcal{S}_p^{ij}|$ is independent of $m$ and it is assumed that $|\mathcal{S}_p^{ij}| \ll m$.

**Corollary 1** *If $D^{ij} = 0$ for $1 \leq i \leq r$, $1 \leq j \leq r$, and $i \neq j$, then the time complexity of* TRADEOFFA1 *is the same as the complexity of* TRADEOFFA.

**Proof:** If $D^{ij} = 0$ for $1 \leq i \leq r$, $1 \leq j \leq r$, and $i \neq j$, then $\pi^{ij} = 1$ and so $\hat{\pi} = 1$. So the time complexity of TRADEOFFA1 is $\mathcal{O}(m)$. $\square$

**Theorem 11** *The time complexity of computing the equilibrium offers for the package deal procedure is $\mathcal{O}(m\hat{\pi}r^3 T(n - \frac{T}{2}))$ where $T = min(2r - 1, n)$.*

**Proof:** Let $a$ denote the agent that offers at $t = 1$ and assume that $n$ is even (the proof for odd $n$ is analogous). We begin with the last time period and then reason backwards. Since $n$ is even and $a$ starts at $t = 1$, it is $b$'s turn to offer in the last time period. For $t = n$, the time taken to find EUB$(i, j, t)$ (for a given $i$ and $j$) is $\mathcal{O}(m)$ (see Equation 10). Hence, the time taken to find EUB$(i, j, t)$ for all possible types of $b$ (i.e., $1 \leq j \leq r$) is $\mathcal{O}(mr)$. Note that, at this stage, EUB$(i, t-1)$ is known for $1 \leq i \leq r$ (see Equation 12).

Now consider the time period $t = n - 1$. Since $n$ is even, it is $a$'s turn to offer at $t = n - 1$. In order to find A$(i, t)$, we first need to find $\psi$ where $\psi = $ OPTA$(i, t)$. From Equation 16 we know that, for a given $i$, the time to find OPTA$(i, t)$ depends on the time taken to find EUA$(i, j, t)$ which in turn depends on the time to find F$^a(i, j, e, t)$ (see Equation 13). The time taken for F$^a(i, j, e, t)$ depends on the time taken for A$(i, j, t)$. For a given $i$ and a given $j$, the time taken to find A$(i, j, t)$ is the time taken by the function TRADEOFFA. Since EUB$(j, t)$ is already known at time $t$, the time taken by TRADEOFFA1 is $\mathcal{O}(m\hat{\pi})$ (see Theorem 10). The time taken to find F$^a(i, j, e, t)$ is therefore $\mathcal{O}(m\hat{\pi})$. Given this, the time to find EUA$(i, j, t)$ (for a given $i$, and $t$) is $\mathcal{O}(m\hat{\pi}r)$. Hence, for a given $i$, the time to find $\psi = $ OPTA$(i, t)$ is $\mathcal{O}(m\hat{\pi}r^2)$. At this stage, EUB$(\psi, t)$ is known (see the last sentence in the first paragraph of this proof). Consequently, for a given $i$, the time to find A$(i, t)$ is $\mathcal{O}(m\hat{\pi}r^2)$. Recall that each agent knows only its own type and not that of its opponent. Hence we need to determine A$(i, t)$ for all possible types of $a$ (i.e., for $1 \leq i \leq r$). This takes $\mathcal{O}(m\hat{\pi}r^3)$ time. Note that at this stage EUA$(i, j, t)$ is known for all possible values of $i$ and all possible values of $j$ (where $1 \leq i \leq r$ and $1 \leq j \leq r$).

Now consider the time period $t = n - 2$ when it is $b$'s turn to offer. For $t = n - 2$ and a given $i$, the time to find OPTB$(i, t)$ is $\mathcal{O}(m\hat{\pi}r^2)$ and so the time to find OPTB$(i, t)$ for all possible types of $b$ (i.e., for $1 \leq i \leq r$) is $\mathcal{O}(m\hat{\pi}r^3)$.

In the same way, the time required to do all the necessary computation for each time period $t < n$ is $\mathcal{O}(m\hat{\pi}r^3)$. Hence, the total time to find the equilibrium offer for the first time period is $\mathcal{O}((n-1)m\hat{\pi}r^3)$. However, as noted previously, an agreement may or may not occur in the first time period. If an agreement does not take place at $t = 1$, then the agents update their beliefs and compute the equilibrium offer for $t = 2$ with the updated beliefs. The time to compute the equilibrium offer for $t = 2$ is $\mathcal{O}((n-2)m\hat{\pi}r^3)$. This process of updating beliefs and finding the equilibrium offer is repeated at most $T = min(2r - 1, n)$ times. Hence the time complexity of the package deal is $\Sigma_{i=1}^{T}\mathcal{O}((n-i)m\hat{\pi}r^3) = \mathcal{O}(m\hat{\pi}r^3 T(n - \frac{T}{2}))$ (see Cormen et al., 2003, – page 47 – for details on how to simplify an expression of the form $\Sigma_{i=1}^{T}\mathcal{O}((n-i)m\hat{\pi}r^3)$). $\square$

**Theorem 12** *The package deal procedure generates a Pareto optimal outcome.*

**Proof:** This follows from Theorem 3. The difference between the complete information setting of Theorem 3 and the current incomplete information setting is that for the former setting the agents maximise their cumulative utilities, whereas in the current setting they maximise their expected cumulative utilities. Specifically, for every time period, the offering agent maximises its expected cumulative utility from all the $m$ issues such that its opponent's expected cumulative utility is equal to what the opponent would get from its own equilibrium offer for the next time period. Hence, for the current setting, the equilibrium offer for every time period is Pareto optimal. $\square$

## 4.2 The Simultaneous Procedure

Recall that for this procedure, the $\mu > 1$ partitions are discussed in parallel but independently of each other. The offers made during the negotiation for any one partition do not affect the offers for the others. Specifically, negotiation for each partition starts at $t = 1$ and each partition is settled using the package deal procedure. Since each partition is dealt with separately, the results of Theorem 8 apply directly to each of the $\mu$ partitions.

Let $\hat{\pi}_c$ denote $\hat{\pi}$ for the $c$th partition. Then, from Theorem 11, we know that the time taken for the $c$th (for $1 \le c \le \mu$) partition is $\mathcal{O}(|S_c|\hat{\pi}_c r^3 T(n - \frac{T}{2}))$. Let the partition for which $|S_c|\hat{\pi}_c$ is highest be denoted $S_z$. Then the time complexity of the simultaneous procedure is $\mathcal{O}(|S_z|\hat{\pi}_z r^3 T(n - \frac{T}{2}))$. Also, from Theorem 5, it follows that the simultaneous procedure may not generate a Pareto optimal outcome. Finally, from Theorem 9 we know that the simultaneous procedure has a unique equilibrium outcome if the following condition is satisfied:

$C_5$. If there is no partition $c$ (where $1 \le c \le \mu$) for which the condition $(\neg C_3 \lor C_4)$ is false.

## 4.3 The Sequential Procedure

For this procedure, the $\mu > 1$ partitions are discussed independently and one after another. Also, for $1 \le c \le \mu$, negotiation on the $c$th partition starts in the time period that follows an agreement on the $(c-1)$th partition. Since the package deal is used for each partition, the following results are obtained on the basis of Theorem 8.

First, Theorem 8 applies to each of the $\mu > 1$ partitions. Thus, for the sequential procedure, if negotiation for the $c$th (for $1 \le c \le \mu$) partition starts at time $t_c$, then it ends at the earliest at time $t_c$ and at the latest by $t_c + min(2r - 1, n)$. Second, it follows from Theorem 11 that the time taken for the sequential procedure is $\mathcal{O}(|S_z|\hat{\pi}_z r^3 T(n - \frac{T}{2}))$. Third, the sequential procedure may not generate a Pareto optimal outcome (see Theorem 5). Finally, the conditions for uniqueness are the same as those for the simultaneous procedure.

## 4.4 The Optimal Procedure

Having obtained the equilibrium outcomes for the three procedures for the above defined incomplete information scenario, we now compare them in terms of the expected utilities they generate to each player. Again, the procedure that gives a player the maximum expected utility is the optimal one.

**Theorem 13** *The package deal is optimal for each agent.*

**Proof:** The proof for this is the same as Theorem 7. The only difference between the complete information setting of Theorem 7 and the current incomplete information setting is that for the package deal procedure for the former setting (during time period $t < n$), the offering agent proposes a package that maximises its own cumulative utility, while giving its opponent a cumulative utility equal to what the opponent would get from its own equilibrium offer in the next time period. On the other hand, for the current incomplete information setting, the offering agent proposes a package that maximises its own expected cumulative utility while giving its opponent an expected cumulative utility equal to what the opponent would get from its own equilibrium offer in the next time period. Also, for each agent, the package deal maximises the expected cumulative utility from all the $m$ issues (since tradeoffs are made across all the $m$ issues). But the simultaneous procedure maximises each agent's expected cumulative utility for each partition (i.e., the simultaneous procedure does not make tradeoffs across partitions). Hence each agent's expected cumulative utility for all the $m$ issues is higher for the package deal relative to the simultaneous procedure. Furthermore, irrespective of how the $m$ issues are partitioned into $\mu$ partitions, we know that the simultaneous procedure is better than the sequential one for each agent (see Theorem 7). Hence, the package deal is optimal for each agent. □

| | Package deal | Simultaneous | Sequential |
|---|---|---|---|
| Time of agreement | Earliest: 1<br>Latest: $min(2r-1, n)$<br>for all $m$ issues | Earliest: 1<br>Latest: $min(2r-1, n)$<br>for all $m$ issues | For the $c$th partition<br>$t_c^e = t_c^s$<br>$t_c^l = t_c^s + min(2r-1, n)$<br>for $1 \le c \le \mu$ |
| Time to compute equilibrium | $\mathcal{O}(m\hat{\pi}r^3T(n-\frac{T}{2}))$ | $\mathcal{O}(|S_z|\hat{\pi}_z r^3T(n-\frac{T}{2}))$ | $\mathcal{O}(|S_z|\hat{\pi}_z r^3T(n-\frac{T}{2}))$ |
| Pareto optimal? | Yes | No | No |
| Unique equilibrium? | If $\neg C_3 \lor C_4$ | If $C_5$ | If $C_5$ |

Table 3: A comparison of the expected outcomes for the three multi-issue procedures for the symmetric information setting (for the sequential procedure, $t_c^s$ denotes the start time for the $c$th partition, $t_c^e$ the earliest possible time of agreement, and $t_c^l$ the latest possible time of agreement).

These results are summarised in Table 3.

## 5. Multi-Issue Negotiation with Asymmetric Uncertainty about the Opponent's Utility

In some bargaining situations, one of the players may know something of relevance that the other may not know. For example, when bargaining over the price of a second hand car, the seller knows its quality but the buyer does not. Such situations are said to have *asymmetry* in information between the players (Muthoo, 1999). Our asymmetric information setting differs from the symmetric one explored in the previous section in that one of the two agents (say $a$) has complete information, but the other (say $b$) is uncertain about $a$'s utility function: for $1 \le c \le m$, agent $b$ is uncertain about $k_c^a$. Here, $K$, $P^a$, $P^b$, $n$, $r$, and $m$ are as defined in Section 4. The negotiation parameters $K$, $P^a$, $P^b$, $r$, $\delta$, $n$, and $m$ are common knowledge to the negotiators. Furthermore, $a$ knows its own type and that of $b$, while $b$ knows its own type but not that of $a$. Finally, the definitions for the cumulative utility functions remain the same as in Section 4. For this setting, we now determine the equilibrium for each of the three multi-issue procedures.

## 5.1 The Package Deal Procedure

We extend the analysis of Section 4 to the current setting as follows. It is clear that for the last time period ($t = n$), the utilities $\text{EUA}(i, t)$ and $\text{EUB}(i, t)$ are as per Section 4. Let $\bar{j}$ denote $b$'s actual type. Recall that agent $a$ now knows $\bar{j}$. Hence on the basis of Equation 13 for the $SU_I$ setting, we get $\text{EUA}(i, j, t)$ for the current asymmetric information setting as follows:

$$\text{EUA}(i, j, t) \;=\; F^a(i, j, \bar{j}, t) \quad \text{for } 1 \le i \le r \text{ and } 1 \le j \le r \tag{19}$$

On the other hand, since agent $b$ is uncertain about $a$'s type, the definitions for $\text{EUB}(i, t)$ and $\text{EUB}(i, j, t)$ are as given in Section 4. Also, the definitions for $F^a$, $F^b$, $\text{A}(i, j, t)$, $\text{B}(i, j, t)$, $\text{OPTA}(i, t)$, and $\text{OPTB}(i, t)$ for all time periods remain the same as in Section 4.

Finally, in this setting, belief updating does not apply to agent $a$ because it has complete information. Only agent $b$ updates its beliefs about $a$. This is done in the same way described in Section 4. Because of $b$'s uncertainty, we use the concept of sequential equilibrium in this setting as well. The following theorem characterises the equilibrium for the package deal procedure.

**Theorem 14** *For the package deal procedure the following strategies form a sequential equilibrium. The equilibrium strategies for $t = n$ are:*

$$\text{A}(i, n) = \begin{cases} \textit{OFFER} \;\; [\delta^{n-1}, \boldsymbol{0}] & \textit{IF a's TURN} \\ \textit{ACCEPT} & \textit{IF b's TURN} \end{cases}$$

$$\text{B}(i, n) = \begin{cases} \textit{OFFER} \;\; [\boldsymbol{0}, \delta^{n-1}] & \textit{IF b's TURN} \\ \textit{ACCEPT} & \textit{IF a's TURN} \end{cases}$$

*for $1 \le i \le r$. For all preceding time periods $t < n$, if $[x^t, y^t]$ denotes the offer made at time $t$, then the equilibrium strategies are defined as follows:*

$$\text{A}(i, t) = \begin{cases} \textit{OFFER} \;\; \textsc{tradeoffa1}(K, \delta, \text{EUB}(\bar{j}, t), i, \bar{j}, m, t, P^a, P^b) & \textit{IF a's TURN} \\ \textit{RECEIVE OFFER} & \textit{IF b's TURN} \\ \textit{If } (U_i^a([x^t, y^t], t) \ge \text{EUA}(i, t)) \textit{ ACCEPT else REJECT} \end{cases}$$

$$\text{B}(i, t) = \begin{cases} \textit{OFFER} \;\; \textsc{tradeoffb1}(K, \delta, \text{EUA}(\phi, t), i, \phi, m, t, P^a, P^b) & \textit{IF b's TURN} \\ \textit{If offer gets rejected UPDATE BELIEFS} \\ \textit{RECEIVE OFFER and UPDATE BELIEFS} & \textit{IF a's TURN} \\ \textit{If } (U_i^b(x^t, y^t], t) \ge \text{EUB}(i, t)) \textit{ ACCEPT else REJECT} \end{cases}$$

*for $1 \le i \le r$. Here, $\bar{j}$ denotes agent $b$'s type and $\phi = \text{OPTB}(i, t)$. The earliest possible time of agreement is $t = 1$ and the latest possible time is $t = min(2r - 1, n)$.*

**Proof:** As Theorem 8. The only difference is that $a$ now knows $b$'s type ($\bar{j}$). Hence this information is used as a parameter for $\textsc{tradeoffa1}$.

The earliest possible time of agreement is $t = 1$. We show this with the following example. Let $n = 2$, $m = 2$, $r = 2$, $\delta = 1/2$, and $K = [1, 2; 5, 1]$. Let $b$ (i.e., the agent with uncertain information) be the offering agent at time $t = 1$. Assume that $b$ is of type 2 (i.e., $k^b = [5, 1]$). Let $P^a(1) = 0.9$ and $P^a(2) = 0.1$. Since $r = 2$, $b$ can play two possible strategies at time $t = 1$: one that corresponds to the case where $a$ is of type 1 and the other that corresponds to the case where $a$ is of type 2. For the former case, $b$'s equilibrium offer at $t = 1$ is $[0, 1]$ for the first issue and $[\frac{3}{4}, \frac{1}{4}]$ for the second. Hence $\text{EUB}(1, 1, 1) = 4.725$. For the latter case, $b$'s equilibrium offer at $t = 1$ is $[\frac{2}{5}, \frac{3}{5}]$ for the first issue and $[1, 0]$ for the second one. Hence $\text{EUB}(1, 2, 1) = 3$. Since $\text{EUB}(1, 1, 1) > \text{EUB}(1, 2, 1)$, $\text{OPTB}(1, 1) = 1$ and $b$ plays the former strategy. Now if $a$ is in fact of type 1, then it accepts $b$'s offer at $t = 1$. But if $a$ is in fact of type 2, it rejects $b$'s offer at $t = 1$ since it can get a higher utility at $t = 2$. An agreement therefore occurs at $t = 2$. Thus, the earliest possible time of agreement is $t = 1$.

Now consider the case where an agent $b$ of type $i$ offers at $t = 1$ but an agreement does not occur at this time. When $b$'s offer gets rejected, it knows that $a$ is not of type $\text{OPTB}(i, 1)$. Thus the number of possible types for $a$ is now reduced to $r - 1$. This happens every time $b$ makes an offer (i.e., every alternate time period) but it gets rejected. When negotiation reaches time period $t = 2r - 1$, there is only one possible type for $a$. Since $a$ knows $b$'s type, an agreement therefore takes place at $t = 2r - 1$. However, if $n < 2r - 1$ then an agreement occurs at $t = n$ (see $\text{A}(i, n)$ and $\text{B}(i, n)$). In other words, if an agreement does not occur at $t = 1$, then it occurs at the latest by $t = min(2r - 1, n)$. $\square$

Note that the latest possible time of agreement for the asymmetric information setting is the same as that for the symmetric information setting of Theorem 8. This is because, in the asymmetric setting, although $a$ knows $b$'s type, $b$ is uncertain about $a$'s type. Also, it takes $2r - 1$ time periods for $b$ to come to know $a$'s actual type. Hence, the earliest and latest time of agreement is the same for both settings.

**Theorem 15** *The time complexity of computing the equilibrium offers for the package deal procedure is $\mathcal{O}(m\hat{\pi}r^3\frac{T}{2}(n - \frac{T}{2}))$ where $T = min(2r - 1, n)$.*

**Proof:** Let $a$ denote the agent that offers at $t = 1$ and assume that $n$ is even (the proof for odd $n$ is analogous). We begin with the last time period and then reason backwards. Since $n$ is even and agent $a$ starts at $t = 1$, it is $b$'s turn to offer in the last time period. For $t = n$, the time taken to find $\text{EUB}(i, j, t)$ (for a given $i$ and $j$) is $\mathcal{O}(m)$ (see Equation 10). Hence, the time taken to find $\text{EUB}(i, j, t)$ for all possible types of $b$ (i.e., $1 \leq j \leq r$) is $\mathcal{O}(mr)$. Note that, at this stage, $\text{EUB}(i, t - 1)$ is known for $1 \leq i \leq r$ (see Equation 12).

Now consider the time period $t = n - 1$. Since $n$ is even, it is $a$'s turn to offer at $t = n - 1$. In order to find $\text{A}(i, t)$, we first need to find $\psi$ where $\psi = \text{OPTA}(i, t)$. From Equation 16 we know that, for a given $i$, the time to find $\text{OPTA}(i, t)$ depends on the time taken to find $\text{EUA}(i, j, t)$ which, in turn, depends on the time to find $\text{F}^a(i, j, e, t)$ (see Equation 19). The time taken for $\text{F}^a(i, j, e, t)$ depends on the time taken for $\text{A}(i, j, t)$. For a given $i$ and a given $j$, the time taken to find $\text{A}(i, j, t)$ is the time taken by $\text{TRADEOFFA1}$. Since $\text{EUB}(j, t)$ is already known at time $t$, the time taken by the function $\text{TRADEOFFA1}$ is $\mathcal{O}(m\hat{\pi})$ (as Theorem 2). The time taken to find $\text{F}^a(i, j, e, t)$ is therefore $\mathcal{O}(m\hat{\pi})$. Given this, the time to find $\text{EUA}(i, j, t)$ (for a given $i$, $j$, and $t$) is $\mathcal{O}(m\hat{\pi})$ since $b$'s type is known to both agents – see Equation 19. Hence, for a given $i$, the time to find $\psi = \text{OPTA}(i, t)$ is $\mathcal{O}(m\hat{\pi}r)$. At this stage, $\text{EUB}(\psi, t)$ is known (see the last sentence in the first paragraph of this proof). Consequently, for a given $i$, the time to find $\text{A}(i, t)$ is $\mathcal{O}(m\hat{\pi}r)$. Recall that $b$ does not know $a$'s type. Hence we need to determine $\text{A}(i, t)$ for all possible types of $a$ (i.e., for $1 \leq i \leq r$). This takes $\mathcal{O}(m\hat{\pi}r^2)$ time. Note that at this stage $\text{EUA}(i, j, t)$ is known for all possible values of $i$ and all possible values of $j$ (where $1 \leq i \leq r$ and $1 \leq j \leq r$).

Now consider the time period $t = n - 2$ when it is $b$'s turn to offer. The only difference between the computation for $t = n - 1$ and $t = n - 2$ is that for the former case, the time to find $\text{EUA}(i, j, t)$ (for a given $i$, $j$, and $t$) is $\mathcal{O}(m\hat{\pi})$ since $b$'s type is known to both agents. However for the latter case, the time to find EUB$(i, j, t)$ (for a given $i$, $j$, and $t$) is $\mathcal{O}(m\hat{\pi}r)$ since $a$'s type is not known to $b$ (see Equation 14). Consequently, for a given $i$, the time to find B$(i, t)$ is $\mathcal{O}(m\hat{\pi}r^2)$. So the time to determine B$(i, t)$ for all possible types of $b$ (i.e., for $1 \leq i \leq r$) is $\mathcal{O}(m\hat{\pi}r^3)$ time. Note that at this stage EUB$(i, j, t)$ is known for all possible values of $i$ and all possible values of $j$ (where $1 \leq i \leq r$ and $1 \leq j \leq r$).

In the same way, the time required to do all the necessary computation for each odd time period $t < n$ is $\mathcal{O}(m\hat{\pi}r^2)$, while that for each even time period is $\mathcal{O}(m\hat{\pi}r^3)$. Hence, the total time to find the equilibrium offer for the first time period is $\mathcal{O}(m\hat{\pi}r^3(\frac{n-1}{2}))$. However, as noted previously, an agreement may or may not occur in the first time period. If an agreement does not take place at $t = 1$, then the agents update their beliefs and compute the equilibrium offer for $t = 2$ with the updated beliefs. The time to compute the equilibrium offer for $t = 2$ is $\mathcal{O}(m\hat{\pi}r^3(\frac{n-2}{2}))$. This process of updating beliefs and finding the equilibrium offer is repeated at most $T = min(2r - 1, n)$ times. Hence the time complexity of the package deal is $\Sigma_{i=1}^{T}\mathcal{O}(m\hat{\pi}r^3(\frac{n-i}{2})) = \mathcal{O}(m\hat{\pi}r^3(n - \frac{T}{2})\frac{T}{2})$. $\square$

**Theorem 16** *The package deal procedure generates a Pareto optimal outcome.*

**Proof:** As per Theorem 12. $\square$

**Theorem 17** *For a given first mover, the package deal procedure has a unique equilibrium outcome if $\neg C_3 \vee C_4$ is true.*

**Proof:** As per Theorem 9. $\square$

### 5.2 The Simultaneous Procedure

Theorem 14 applies to each of the $\mu > 1$ partitions. Hence, from Theorem 15, we know that the time taken for the $c$th (for $1 \leq c \leq \mu$) partition is $\mathcal{O}(|S_c|\hat{\pi}_cr^3(\frac{n-T}{2})\frac{T}{2})$. Hence, the time complexity of the simultaneous procedure is $\mathcal{O}(|S_z|\hat{\pi}_zr^3(n - \frac{T}{2})\frac{T}{2})$. Also, from Theorem 5, it follows that the simultaneous procedure may not generate a Pareto optimal outcome. Finally, from Theorem 17 we know that the simultaneous procedure has a unique equilibrium outcome if the condition $C_5$ is true.

### 5.3 The Sequential Procedure

First, Theorem 14 applies to each of the $\mu > 1$ partitions. Thus, for the sequential procedure, if negotiation for the $c$th (for $1 \leq c \leq \mu$) partition starts at time $t_c$, then it ends at the earliest at time $t_c$ and at the latest by $t_c + min(2r - 1, n)$. Second, it follows from Theorem 15 that the time taken for the sequential procedure is $\mathcal{O}(|S_z|\hat{\pi}_zr^3(n - \frac{T}{2})\frac{T}{2})$. Third, the sequential procedure may not generate a Pareto optimal outcome (see Theorem 5). Finally, the conditions for uniqueness are the same as those for the simultaneous procedure.

### 5.4 The Optimal Procedure

It follows from Theorem 13 that, for each agent, the optimal procedure is the package deal. These results are summarised in Table 4.

## 6. Multi-Issue Negotiation for Interdependent Issues

For the independent issues case of Section 4, an agent's utility for issue $c$ (for $1 \leq c \leq m$) depends only on its share for that issue and is independent of other issues. However, in many cases, an agent's utility from an issue depends not only on its share for the issue, but also on its share for others (Klein et al., 2003). Given this, in this section we focus on such interdependent issues. Specifically, we model interdependence between the issues as follows. Consider a package $[x^t, y^t]$. For this package, for an agent $a$ of type $i$, the utility from issue $c$ at time $t$ is now of the form:

$$u_{ic}^a([x^t, y^t], t) = \begin{cases} K_{ic}x_c + \Sigma_{j=1}^m \chi_{ij}(x_c - x_j) & \text{if } t \le n \\ 0 & \text{otherwise} \end{cases} \tag{20}$$

and that for an agent $b$ of type $i$, it is:

$$u_{ic}^b([x^t, y^t], t) = \begin{cases} K_{ic}y_c + \Sigma_{j=1}^m \chi_{ij}(y_c - y_j) & \text{if } t \le n \\ 0 & \text{otherwise} \end{cases} \tag{21}$$

where $K_{ic}$ denotes a constant positive real number and $\chi_{ij}$ a constant real number that may be either positive or negative. As before, an agent's cumulative utility is the sum of its utilities from the individual issues:

$$U_i^a([x^t, y^t], t) = \begin{cases} \Sigma_{c=1}^m \bar{K}_{ic}x_c^t & \text{if } t \le n \\ 0 & \text{otherwise} \end{cases} \tag{22}$$

$$U_i^b([x^t, y^t], t) = \begin{cases} \Sigma_{c=1}^m \bar{K}_{ic}y_c^t & \text{if } t \le n \\ 0 & \text{otherwise} \end{cases} \tag{23}$$

|  | Package deal | Simultaneous | Sequential |
|---|---|---|---|
| Time of agreement | Earliest: 1<br>Latest: $min(2r-1, n)$<br>for all $m$ issues | Earliest: 1<br>Latest: $min(2r-1, n)$<br>for all $m$ issues | For the $c$th partition<br>$t_c^e = t_c^s$<br>$t_c^l = t_c^s + min(2r-1, n)$<br>for $1 \le c \le \mu$ |
| Time to compute equilibrium | $\mathcal{O}(m\hat{\pi}r^3\frac{T}{2}(n-\frac{T}{2}))$ | $\mathcal{O}(\|S_z\|\hat{\pi}_z r^3(n-\frac{T}{2})\frac{T}{2})$ | $\mathcal{O}(\|S_z\|\hat{\pi}_z r^3(n-\frac{T}{2})\frac{T}{2})$ |
| Pareto optimal? | Yes | No | No |
| Unique equilibrium? | If $\neg C_3 \lor C_4$ | If $C_5$ | If $C_5$ |

Table 4: A comparison of the expected outcomes for the three multi-issue procedures for the asymmetric information setting (for the sequential procedure, $t_c^s$ denotes the start time for the $c$th partition, $t_c^e$ the earliest possible time of agreement, and $t_c^l$ the latest possible time of agreement).

Here $\bar{K}$ denotes a vector analogous to the vector $K$ except that the individual elements of the latter are all constant positive real numbers, while those of the former may be positive or negative. Note that in Equations 5 and 6, all the coefficients are positive (i.e., $K_{ic} > 0$ for $1 \le i \le r$ and $1 \le c \le m$). But in Equations 22 and 23, the coefficient ($\bar{K}_{ic}$) may be a positive or a negative real number.

The above cumulative utility functions are linear (see Pollak, 1976; Charness & Rabin, 2002; Sobel, 2005, for other forms of utility functions for interdependent preferences[10]). As mentioned before, we chose the linear form for reasons of computational tractability.

In this setting the vector $\bar{K}$ and the functions $P^a$ and $P^b$ are common knowledge to the negotiators. Also, each agent knows its own type, but not that of its opponent. In addition, each agent knows $r$, $\delta$, $n$, and $m$. In other words, there is symmetric uncertainty about the opponent's utility (as we will see in Section 6.4, the results for the asymmetric case can easily be obtained from the following analysis for the symmetric case).

## 6.1 The Package Deal Procedure

For the cumulative utilities defined in Equations 22 and 23, Theorem 18 characterises the equilibrium for the package deal.

**Theorem 18** *For the package deal procedure, the following strategies form a sequential equilibrium. The equilibrium strategies for $t = n$ are:*

$$\text{A}(i, n) = \left\{ \begin{array}{ll} \textit{OFFER } [\delta^{n-1}, \boldsymbol{0}] & \textit{IF a's TURN} \\ \textit{ACCEPT} & \textit{IF b's TURN} \end{array} \right.$$

$$\text{B}(i, n) = \left\{ \begin{array}{ll} \textit{OFFER } [\boldsymbol{0}, \delta^{n-1}] & \textit{IF b's TURN} \\ \textit{ACCEPT} & \textit{IF a's TURN} \end{array} \right.$$

*for $1 \leq i \leq r$. For all preceding time periods $t < n$, if $[x^t, y^t]$ denotes the offer made at time $t$, then the equilibrium strategies are defined as follows:*

$$\text{A}(i, t) = \left\{ \begin{array}{ll} \textit{OFFER } \text{TRADEOFFA1}(\bar{K}, \delta, \text{EUB}(\psi, t), i, \psi, m, t, P^a, P^b) & \textit{IF a's TURN} \\ \textit{If offer gets rejected UPDATE BELIEFS} & \\ \textit{RECEIVE OFFER and UPDATE BELIEFS} & \textit{IF b's TURN} \\ \textit{If } (U_i^a([x^t, y^t], t) \geq \text{EUA}(i, t)) \textit{ ACCEPT else REJECT} & \end{array} \right.$$

$$\text{B}(i, t) = \left\{ \begin{array}{ll} \textit{OFFER } \text{TRADEOFFB1}(\bar{K}, \delta, \text{EUA}(\phi, t), i, \phi, m, t, P^a, P^b) & \textit{IF b's TURN} \\ \textit{If offer gets rejected UPDATE BELIEFS} & \\ \textit{RECEIVE OFFER and UPDATE BELIEFS} & \textit{IF a's TURN} \\ \textit{If } (U_i^b(x^t, y^t], t) \geq \text{EUB}(i, t)) \textit{ ACCEPT else REJECT} & \end{array} \right.$$

*for $1 \leq i \leq r$. Here, $\psi = \text{OPTA}(i, t)$ and $\phi = \text{OPTB}(i, t)$. The earliest possible time of agreement is $t = 1$ and the latest possible time is $t = min(2r - 1, n)$.*

**Proof:** As Theorem 8. The only difference between the independent issues setting of Theorem 8 and the present interdependent issues one is in terms of the definition for cumulative utilities: in Equations 5 and 6, all the coefficients are positive (i.e., $K_{ic} > 0$ for $1 \leq i \leq r$ and $1 \leq c \leq m$). But in Equations 22 and 23, the coefficient ($\bar{K}_{ic}$) may be a positive or a negative real number. However, the greedy method (given in Theorem 1) for solving the fractional knapsack problem of Equation 15 works for both positive and negative coefficients (Martello & Toth, 1990; Cormen et al., 2003). Hence, the proof of Theorem 8 applies to this setting as well. $\square$

---

10. Although in (Pollak, 1976; Charness & Rabin, 2002; Sobel, 2005) these forms are discussed in the context of how an agent's utility depends on the utility of other agent's, they may equally well be interpreted for the case where an agent's utility for an issue depends on its share for other issues.

**Theorem 19** *The time complexity of computing the equilibrium offers for the package deal proce-dure is* $\mathcal{O}(m\hat{\pi}r^3T(n-\frac{T}{2}))$ *where* $T = min(2r-1, n)$.

**Proof:** As Theorem 11. Since the method for making tradeoffs is the same as that for the setting with symmetric uncertainty and independent issues (i.e., $SU_I$), the time complexity is the same as in Theorem 11. $\square$

It is obvious that Theorems 9 and 12 extend to this setting as well.

### 6.2 The Simultaneous Procedure

It follows from above that all the results of Section 4.2 apply to this setting as well.

### 6.3 The Sequential Procedure

It also follows from above that the results of Section 4.3 apply to this setting as well.

### 6.4 The Optimal Procedure

It follows from Theorem 13 that the package deal remains the optimal procedure even if the issues are interdependent. The results for this setting are the same as those in Section 4 and are summarised in Table 3.

Finally, consider the asymmetric information setting of Section 5 but in the current context of interdependent issues. From the above analysis for symmetric uncertainty with interdependent issues, it is clear that the method for making tradeoffs remains the same irrespective of whether the information is symmetric or asymmetric. Consequently, for the case of asymmetric information with interdependent issues, we get the same results as those in Section 5.

Recall that this analysis was done for linear cumulative utilities. We now discuss how our results would hold for more complex utility functions that are non-linear[11]. For cumulative utilities that are nonlinear, the tradeoff problem becomes a *global optimization problem* with a nonlinear objective function. Due to their computational complexity, such nonlinear optimization problems can only be solved using *approximation methods* (Horst & Tuy, 1996; Bar-Yam, 1997; Klein et al., 2003). In contrast, our tradeoff problem is a linear optimization problem, the *exact* solution to which can be found in polynomial time (as shown in Theorems 1 and 2). Although our results apply to linear cumulative utilities, it is not difficult to see how they would hold for the nonlinear case. First, the time of agreement for our case would hold for other (nonlinear) functions. This is because this time depends not on the actual definition of the agents' cumulative utilities but on the information setting (i.e., whether or not the information is complete). Second, let $\mathcal{O}(\omega)$ denote the time complexity of TRADEOFFA1 for nonlinear utilities for the package deal with $\mu = 1$, and $\mathcal{O}(\omega_c)$ that for the $c$th partition. Also, let $S_z$ denote the partition for which $\mathcal{O}(\omega_z)$ is the highest between all partitions. Then, we know from Theorem 11 that the time complexity of the package deal for the setting with symmetric uncertainty is $\mathcal{O}(\omega r^3 T(n-\frac{T}{2}))$. Consequently, the time complexity of both the simultaneous and the sequential procedures is $\mathcal{O}(\omega_z r^3 T(n-\frac{T}{2}))$. Third, while the package deal outcome for our additive cumulative utilities is Pareto optimal, the package deal outcome for nonlinear utilities may not be Pareto optimal. This is because (as stated above) nonlinear optimization problems can only be solved using approximation methods while the linear optimization problem can be solved using an exact method (as in proof of Theorem 1). Finally, since the conditions for a unique solution depend on the actual definition of cumulative utilities, the conditions given in Tables 1 2, 3, and 4 may not hold for other forms of utility functions.

---

11. Note that bilateral bargaining for which the players' utility functions are nonlinear has been studied by Hoel (1986) in the context of a single issue as opposed to the multi-issue case which is the focus of our study.

## 7. Related Work

Since Schelling (1956) first noted the fact that the outcome of negotiation depends on the choice of negotiation procedure, much research effort has been devoted to the study of different procedures for negotiating multiple issues. For instance, Fershtman (1990) extended the model developed by Rubinstein (1982), for splitting a single pie, to sequential negotiation for two pies. However, this model assumes complete information, imposes an agenda exogenously, and then studies the relation between the agenda and the outcome of the sequential bargaining game. In more detail, for two pies of different sizes, he analyses the effect of going first on the large and the small pie.

A number of researchers have also studied negotiations with an endogenous agenda (Inderst, 2000; In & Serrano, 2003; Bac & Raff, 1996). In Inderst (2000) players have discount factors, but no deadlines. For independent issues, this work assumes complete information and studies three different negotiation procedures: package deal, simultaneous, and sequential negotiation with endogenous agenda. Their main result is that the package deal is the optimal procedure and that for each procedure there exist multiple equilibria. In and Serrano (2003) extend this work by finding conditions under which the equilibrium becomes unique. Note that our work differs from both of these in that we analyse negotiations with both discount factors and deadlines, which we consider to be much more common with automated negotiations. Moreover, we do this for both independent and interdependent issues without making the complete information assumption.

Bac and Raff (1996) also developed a model that has an endogenous agenda. They extended the model developed by Rubinstein (1985) for single pie bargaining with incomplete information by adding a second pie. In this model, the players have discount factors, but no deadlines. The size of the pie is known to both agents and the discounting factor is assumed to be equal for all the issues for both agents. Also, there is asymmetric information: one of the players knows its own discounting factor and that of its opponent, while the other player knows its own discounting factor, but is uncertain of its opponent's. In more detail, this factor can take one of two values, $\delta_H$ with probability $x$, and $\delta_L$ with probability $1 - x$. These probabilities are common knowledge. For this model, the authors determine the equilibrium for the package deal and the sequential procedure. They show that, under certain conditions, the sequential procedure can be the optimal one. However, there are three key differences between this model and ours. First, we analyse both symmetric and asymmetric information settings, while Bac and Raff analyse only the latter. Second, the negotiators in our model have a deadline, while in Bac and Raff they do not. Again, we believe our analysis covers situations that often occur in automated negotiation settings. Finally, Bac and Raff focus on independent issues, but we analyse both independent and interdependent issues.

A slightly different approach (from the above ones) was taken by Busch and Horstmann (1997). Again, they extended the model developed by Rubinstein (1985), but by adding a preliminary period in which the agents bargain over the agenda. The outcome of this stage is then used as the agenda for negotiating over the issues. In this complete information model, there are two pies for bargaining. Furthermore, these two issues become available for negotiation at different time points. The players have discount factors or fixed time costs, but no deadlines. Since there are two issues, there are two possible agendas. The outcome for these two agendas is compared with that for the package deal. Their main result is that the players may have conflicting preferences over the optimal agenda. Note that a key difference between this model and ours is that all the issues in our model are available from the beginning, while in their model the two issues become available at different time points. Furthermore, Busch and Horstman assume complete information, while we do not.

From all the models mentioned above, perhaps the one that is closest to ours is the one developed by Inderst (2000). Unlike our work, Inderst assumes complete information and independent issues. Also, it does not model player deadlines, while we do. However, Inderst does model players' time preferences as discount factors. Also, just like our model, all the issues for negotiation are available at the beginning of negotiation. In terms of results, Inderst shows that the package deal is the optimal procedure. Our study also shows that the package deal is the optimal procedure for both agents. Finally, our work provides a detailed analysis of the attributes of the different procedures (such as the time of agreement, the time complexity, the Pareto optimality, and the conditions for uniqueness), while Inderst does not.

In summary, all the aforementioned models for multi-issue negotiation differ from ours in at least one of three major ways. The players in our model have both discount factors and deadlines, but a general characteristic of the above models is that the players only have discount factors but no deadlines[12]. Negotiation with deadlines has been studied by Sandholm and Vulkan (1999) (in the context of a single issue) and by Fatima et al. (2004) for the sequential procedure with $\mu = m$. Given this, our contribution lies firstly in finding the equilibrium for all the three procedures. Second, we analyse both asymmetric and symmetric information settings, while previous work analyses only the former. Third, we analyse both independent and interdependent issues while previous work focuses primarily on independent issues. Furthermore, the existing literature does not compare the different multi-issue procedures in terms of their attributes (viz. time complexity, Pareto optimality, uniqueness, and time of agreement). By considering these, our study allows a more informed choice to be made about a wider range of tradeoffs that are involved in determining which is the most appropriate procedure.

Finally, we would like to point that in Fatima et al. (2006), we considered independent issues and carried out the same study as we do in this work, but in a symmetric information setting with uncertainty about the negotiation deadline (as opposed to uncertainty over the agents' utility functions that is the focus of this work). The key result of (Fatima et al., 2006) is similar to the result of our current work, namely that the optimal procedure in (Fatima et al., 2006) is the package deal.

## 8. Conclusions and Future Work

This paper studied bilateral multi-issue negotiation between self-interested agents in a wide range of settings. Each player has time constraints in the form of deadlines and discount factors. Specifically, we considered both independent and interdependent issues and studied the three main multi-issue procedures for conducting such negotiations: the package deal, the simultaneous procedure, and the sequential procedure. We determined equilibria for each procedure for two different information settings. In the first, there is symmetric uncertainty about the opponent's utility. In the second, there is asymmetric uncertainty about the opponent's utility. We analysed both settings for the case of independent and interdependent issues. For each setting, we compared the outcomes of the different procedures and showed that the package deal is optimal for each agent. We then compared the three procedures in terms of four attributes: the time complexity of the procedure, the Pareto optimality of the equilibrium solution, the uniqueness of the equilibrium solution, and the time of agreement (see Table 1).

---

12. (Fatima et al., 2004) studies a multi-issue model with deadlines, but it focuses on determining the equilibrium for one specific sequential procedure: the one in which each partition has a single issue.

In more detail, our study shows that the package deal is in fact the optimal procedure for each party. We also showed that although the package deal may be computationally more complex than the other two procedures, it generates Pareto optimal outcomes (unlike the other two procedures), it has similar earliest and latest possible times of agreement as the simultaneous procedure (which is better than the sequential procedure), and that it (like the other two procedures) generates a unique outcome only under certain conditions (which we defined).

There are several interesting directions for extending the current analysis. First, in this work, we modelled the players' time preferences in the form of discount factors which is the most common basis for such analysis. However, existing literature (Busch & Horstman, 1997) shows that the outcome for negotiation with discount factors can differ from the outcome for negotiation with fixed time costs. It will, therefore, be interesting to extend our results to negotiations with fixed time costs. Second, our present work analysed the setting with uncertainty about utility functions. Generalisation of our results to scenarios with other sources of uncertainties such as the agents' discount factors is another direction for future work.

## Acknowledgements

We are grateful to Sarit Kraus for her detailed comments on earlier versions of this paper. We also thank the anonymous referees; their comments helped us to substantially improve the readability and accuracy of the paper.

## Appendix A. Summary of Notation

$a, b$  The two negotiating agents.

$n$  Negotiation deadline for both agents.

$m$  Total number of issues.

$S$  The set of $m$ issues.

$S_c$  A subset of $S$ ($S_c \subseteq S$).

$M$  Number of issues in the largest partition.

$\mu$  Number of partitions for the simultaneous and sequential procedures.

$\delta_c$  Discount factor for issue $c$ (for $1 \leq c \leq m$).

$\delta$  An $m$ element vector that represents the discount factor for the $m$ issues.

$x^t$  An $m$ element vector that denotes $a$'s share for each of the $m$ issues at time $t$.

$y^t$  An $m$ element vector that denotes $b$'s share for each of the $m$ issues at time $t$.

$[x^t, y^t\ ]$  The package offered at time $t$.

$a_c^t$  Agent $a$'s share for issue $c$ in the equilibrium offer for time period $t$.

$b_c^t$  Agent $b$'s share for issue $c$ in the equilibrium offer for time period $t$.

$a^t$  An $m$ element vector that denotes $a$'s share for each of the $m$ issues in equilibrium at time $t$.

$b^t$  An $m$ element vector that denotes $b$'s share for each of the $m$ issues in equilibrium at time $t$.

$[a^t, b^t\ ]$  The equilibrium package offered at time $t$.

$U_i^a$  Cumulative utility function for agent $a$ of type $i$.

$U_i^b$  Cumulative utility function for agent $b$ of type $i$.

$\text{UA}(t)$  Agent $a$'s cumulative utility from the equilibrium offer for time $t$.

$\text{UB}(t)$  Agent $b$'s cumulative utility from the equilibrium offer for time $t$.

$\text{A}(i, j, t)$  Agent $a$'s equilibrium offer for time $t$ if $a$ is of type $i$ assuming $b$ is type $j$.

$\text{B}(i, j, t)$  Agent $b$'s equilibrium offer for time $t$ if $b$ is of type $i$ assuming $a$ is type $j$.

$\text{A}(i, t)$  Equilibrium strategy for an agent $a$ of type $i$ at time $t$.

$\text{B}(i, t)$  Equilibrium strategy for an agent $b$ of type $i$ at time $t$.

$\text{EUA}(i, t)$  Cumulative utility that an agent $a$ of type $i$ expects to get from $b$'s equilibrium offer at time $t$ (i.e., $a$ is the receiving agent and $b$ the offering agent at $t$).

EUB$(i, t)$ Cumulative utility that an agent $b$ of type $i$ expects to get from $a$'s equilibrium offer at time $t$ (i.e., $b$ is the receiving agent and $a$ the offering agent at $t$).

EUA$(i, j, t)$ Agent $a$'s expected cumulative utility from its equilibrium offer for time $t$ if $a$ is type $i$ and assuming that $b$ is type $j$.

EUB$(i, j, t)$ Agent $b$'s expected cumulative utility from its equilibrium offer for time $t$ if $b$ is type $i$ and assuming $a$ is type $j$.

$r$ Number of types for agent $a$ (and also the number of types for agent $b$).

$T_t^a$ Set of possible types for agent $a$ at time $t$.

$T_t^b$ Set of possible types for agent $b$ at time $t$.

$P^a$ The probability distribution function for $k^a$.

$P^b$ The probability distribution function for $k^b$.

$K$ A vector of $r$ vectors each element of which is in turn a vector of $m$ positive reals.

$\mathcal{S}_p^{ij}$ A subset of $S$ ($\mathcal{S}_p^{ij} \subseteq S$ where $i$ denotes $a$'s type and $j$ that of $b$) such that $|\mathcal{S}_p^{ij}| > 1$ and $\forall_{c,d \in \mathcal{S}_p^{ij}} \frac{K_{ic}}{K_{jc}} = \frac{K_{id}}{K_{jd}}$.

TRADEOFFA Agent $a$'s function for making tradeoffs in the complete information setting.

TRADEOFFB Agent $b$'s function for making tradeoffs in the complete information setting.

TRADEOFFA1 Agent $a$'s function for making tradeoffs in the four incomplete information settings: $SU_I, SU_D, AU_I, AU_D$.

TRADEOFFB1 Agent $b$'s function for making tradeoffs in the four incomplete information settings: $SU_I, SU_D, AU_I, AU_D$.

$\hat{\pi}$ Maximum number of packages that TRADEOFFA1 (or TRADEOFFB1) will have to search to find the one that maximises $a$'s (or $b$'s) expected cumulative utility (considering all possible types of $a$ and $b$).

PA$_t^{ij}$ The set of all possible packages that TRADEOFFA1 can return at time $t$ ($i$ denotes $a$'s type and $j$ that of $b$).

PB$_t^{ij}$ The set of all possible packages that TRADEOFFB1 can return at time $t$ ($i$ denotes $a$'s type and $j$ that of $b$).

## References

Bac, M., & Raff, H. (1996). Issue-by-issue negotiations: the role of information and time preference. *Games and Economic Behavior*, *13*, 125–134.

Bar-Yam, Y. (1997). *Dynamics of Complex Systems*. Addison Wesley.

Binmore, K., Osborne, M. J., & Rubinstein, A. (1992). Noncooperative models of bargaining. In Aumann, R. J., & Hart, S. (Eds.), *Handbook of Game theory with Economic Applications*, Vol. 1, pp. 179–225. North-Holland.

Busch, L. A., & Horstman, I. J. (1997). Bargaining frictions, bargaining procedures and implied costs in multiple-issue bargaining. *Economica*, *64*, 669–680.

Charness, G., & Rabin, M. (2002). Understanding social preferences with simple tests. *The Quarterly Journal of Economics*, *117*(3), 817–869.

Cormen, T. H., Leiserson, C. E., Rivest, R. L., & Stein, C. (2003). *An introduction to algorithms*. The MIT Press, Cambridge, Massachusetts.

Faratin, P., Sierra, C., & Jennings, N. R. (2002). Using similarity criteria to make trade-offs in automated negotiations. *Artificial Intelligence Journal*, *142*(2), 205–237.

Fatima, S. S., Wooldridge, M., & Jennings, N. R. (2002). The influence of information on negotiation equilibrium. In *Agent Mediated Electronic Commerce IV, Designing Mechanisms and Systems*, No. 2531 in LNCS, pp. 180 – 193. Springer Verlag.

Fatima, S. S., Wooldridge, M., & Jennings, N. R. (2004). An agenda based framework for multi-issue negotiation. *Artificial Intelligence Journal*, *152*(1), 1–45.

Fatima, S. S., Wooldridge, M., & Jennings, N. R. (2006). On efficient procedures for multi-issue negotiation. In *Proceedings of the Eighth International Workshop on Agent Mediated Electronic Commerce (AMEC)*, pp. 71–85, Hakodate, Japan.

Fershtman, C. (1990). The importance of the agenda in bargaining. *Games and Economic Behavior*, *2*, 224–238.

Fershtman, C. (2000). A note on multi-issue two-sided bargaining: bilateral procedures. *Games and Economic Behavior*, *30*, 216–227.

Fershtman, C., & Seidmann, D. J. (1993). Deadline effects and inefficient delay in bargaining with endogenous commitment. *Journal of Economic Theory*, *60*(2), 306–321.

Fishburn, P. C. (1988). Normative thoeries of decision making under risk and uncertainty. In Bell, D. E., Raiffa, H., & Tversky, A. (Eds.), *Decision making: Descriptive, normative, and prescriptive interactions*. Cambridge University Press.

Fisher, R., & Ury, W. (1981). *Getting to yes: Negotiating agreement without giving in*. Houghton Mifflin, Boston.

Fudenberg, D., Levine, D., & Tirole, J. (1985). Infinite horizon models of bargaining with one sided incomplete information. In Roth, A. (Ed.), *Game Theoretic Models of Bargaining*. University of Cambridge Press, Cambridge.

Fudenberg, D., & Tirole, J. (1983). Sequential bargaining with incomplete information. *Review of Economic Studies*, *50*, 221–247.

Harsanyi, J. C. (1977). *Rational behavior and bargaining equilibrium in games and social situations*. Cambridge University Press.

Harsanyi, J. C., & Selten, R. (1972). A generalized Nash solution for two-person bargaining games with incomplete information. *Management Science*, *18*(5), 80–106.

Hoel, M. (1986). Perfect equilibria in sequential bargaining games with nonlinear utility functions. *Scandinavian Journal of Economics*, *88*(2), 383–400.

Horst, R., & Tuy, H. (1996). *Global optimazation: Deterministic approaches*. Springer.

In, Y., & Serrano, R. (2003). Agenda restrictions in multi-issue bargaining (ii): unrestricted agendas. *Economics Letters*, *79*, 325–331.

Inderst, R. (2000). Multi-issue bargaining with endogenous agenda. *Games and Economic Behavior*, *30*, 64–82.

Keeney, R., & Raiffa, H. (1976). *Decisions with Multiple Objectives: Preferences and Value Tradeoffs*. New York: John Wiley.

Klein, M., Faratin, P., Sayama, H., & Bar-Yam, Y. (2003). Negotiating complex contracts. *IEEE Intelligent Systems*, *8*(6), 32–38.

Kraus, S. (2001). *Strategic negotiation in multi-agent environments*. The MIT Press, Cambridge, Massachusetts.

Kraus, S., Wilkenfeld, J., & Zlotkin, G. (1995). Negotiation under time constraints. *Artificial Intelligence Journal*, *75*(2), 297–345.

Kreps, D. M., & Wilson, R. (1982). Sequential equilibrium. *Econometrica*, *50*, 863–894.

Lax, D. A., & Sebenius, J. K. (1986). *The manager as negotiator: Bargaining for cooperation and competitive gain*. The Free Press, New York.

Livne, Z. A. (1979). *The role of time in negotiation*. Ph.D. thesis, Massachusetts Institute of Technology.

Lomuscio, A., Wooldridge, M., & Jennings, N. R. (2003). A classification scheme for negotiation in electronic commerce. *International Journal of Group Decision and Negotiation*, *12*(1), 31–56.

Ma, C. A., & Manove, M. (1993). Bargaining with deadlines and imperfect player control. *Econometrica*, *61*, 1313–1339.

Maes, P., Guttman, R., & Moukas, A. (1999). Agents that buy and sell. *Communications of the ACM*, *42*(3), 81–91.

Martello, S., & Toth, P. (1990). *Knapsack problems: Algorithms and computer implementations*. John Wiley and Sons. Chapter 2.

Mas-Colell, A., Whinston, M. D., & Green, J. R. (1995). *Microeconomic Theory*. Oxford University Press.

Muthoo, A. (1999). *Bargaining Theory with Applications*. Cambridge University Press.

Neumann, J. V., & Morgenstern, O. (1947). *Theory of Games and Economic Behavior*. Princeton: Princeton University Press.

Osborne, M. J., & Rubinstein, A. (1990). *Bargaining and Markets*. Academic Press, San Diego, California.

Osborne, M. J., & Rubinstein, A. (1994). *A Course in Game Theory*. The MIT Press.

Pollak, R. A. (1976). Interdependent preferences. *American Economic Review*, *66*(3), 309–320.

Pruitt, D. G. (1981). *Negotiation Behavior*. Academic Press.

Raiffa, H. (1982). *The Art and Science of Negotiation*. Harvard University Press, Cambridge, USA.

Rosenschein, J. S., & Zlotkin, G. (1994). *Rules of Encounter*. MIT Press.

Rubinstein, A. (1982). Perfect equilibrium in a bargaining model. *Econometrica*, *50*(1), 97–109.

Rubinstein, A. (1985). A bargaining model with incomplete information about time preferences. *Econometrica*, *53*, 1151–1172.

Sandholm, T. (2000). Agents in electronic commerce: component technologies for automated negotiation and coalition formation.. *Autonomous Agents and Multi-Agent Systems*, *3*(1), 73–96.

Sandholm, T., & Vulkan, N. (1999). Bargaining with deadlines. In *AAAI-99*, pp. 44–51, Orlando, FL.

Schelling, T. C. (1956). An essay on bargaining. *American Economic Review*, *46*, 281–306.

Schelling, T. C. (1960). *The strategy of conflict*. Oxford University Press.

Sobel, J. (2005). Interdependent preferences and reciprocity. *Journal of Economic Literature*, *XLIII*, 392–436.

Stahl, I. (1972). *Bargaining Theory*. Economics Research Institute, Stockholm School of Economics, Stockholm.

van Damme, E. (1983). *Refinements of the Nash equilibrium concept*. Berlin:Springer-Verlag.

Varian, H. R. (2003). *Intermediate Microeconomics*. W. W. Norton and Company.

Young, O. R. (1975). *Bargaining: Formal theories of negotiation*. Urbana: University of Illinois Press.